\documentclass{elsart}

\usepackage[square,comma]{natbib}
\usepackage{graphicx}
\usepackage{pxfonts}
\usepackage{lineno}

\usepackage{amssymb}

\journal{}

\begin{document}

\thispagestyle{empty}
\begin{Large}
\textbf{DEUTSCHES ELEKTRONEN-SYNCHROTRON}

\textbf{\large{Ein Forschungszentrum der Helmholtz-Gemeinschaft}\\}
\end{Large}

DESY 11-165

September 2011

\begin{eqnarray}
\nonumber &&\cr \nonumber && \cr \nonumber &&\cr
\end{eqnarray}
\begin{eqnarray}
\nonumber
\end{eqnarray}
\begin{center}
\begin{Large}
\textbf{Production of transform-limited X-ray pulses through
self-seeding at the European X-ray FEL}
\end{Large}
\begin{eqnarray}
\nonumber &&\cr \nonumber && \cr
\end{eqnarray}

\begin{large}
Gianluca Geloni,
\end{large}
\textsl{\\European XFEL GmbH, Hamburg}
\begin{large}

Vitali Kocharyan and Evgeni Saldin
\end{large}
\textsl{\\Deutsches Elektronen-Synchrotron DESY, Hamburg}
\begin{eqnarray}
\nonumber
\end{eqnarray}
\begin{eqnarray}
\nonumber
\end{eqnarray}
ISSN 0418-9833
\begin{eqnarray}
\nonumber
\end{eqnarray}
\begin{large}
\textbf{NOTKESTRASSE 85 - 22607 HAMBURG}
\end{large}
\end{center}
\clearpage
\newpage

\begin{frontmatter}



\title{Production of transform-limited X-ray pulses through self-seeding at
the European X-ray FEL}


\author[XFEL]{Gianluca Geloni\thanksref{corr},}
\thanks[corr]{Corresponding Author. E-mail address: gianluca.geloni@xfel.eu}
\author[DESY]{Vitali Kocharyan}
\author[DESY]{and Evgeni Saldin}

\address[XFEL]{European XFEL GmbH, Hamburg, Germany}
\address[DESY]{Deutsches Elektronen-Synchrotron (DESY), Hamburg,
Germany}

\begin{abstract}
An important goal for any advanced X-ray FEL is an option for
providing Fourier-limited X-ray pulses. In this way, no
monochromator is needed in the experimental hall. Self-seeding is a
promising approach to significantly narrow the SASE bandwidth to
produce nearly transform-limited pulses. These are important for
many experiments including 3D diffraction imaging. We discuss the
implementation of a single-crystal self-seeding scheme in the hard
X-ray lines of the European XFEL. For this facility,
transform-limited pulses are particularly valuable since they
naturally support the extraction of more FEL power than at
saturation by exploiting tapering in the tunable-gap baseline
undulators. Tapering consists of a stepwise change of the undulator
gap from segment to segment. Based on start-to-end simulations
dealing with the up-to-date parameters of the European XFEL, we show
that the FEL power reaches about 400 GW, or one order of magnitude
higher power than the SASE saturation level (20 GW). This analysis
indicates that our self-seeding scheme is not significantly affected
by non-ideal electron phase-space distribution, and yields about the
same performance as in the case for an electron beam with ideal
parameters. The self-seeding scheme with a single crystal
monochromator is extremely compact (about 5 m long), and cost
estimations are low enough to consider adding it to the European
XFEL capabilities from the very beginning of the operation phase.
\end{abstract}

%
%
%
\end{frontmatter}



\section{\label{sec:intro} Introduction}

Conventional SASE X-ray FELs like the European X-ray FEL provide
transversely coherent beams, but only limited longitudinal coherence
\cite{LCLS2}-\cite{tdr-2006}. Many experiments, including 3D
diffraction imaging, require both transverse and longitudinal
coherence. In principle, one can create a longitudinally coherent
source by use of a monochromator located in the experimental hall,
but this is often undesirable because of intensity losses. An
important goal for the European X-ray FEL is the production of X-ray
pulses with the minimum allowed photon energy width for a given
pulse length, that is transform-limited pulses.

\begin{figure}[tb]
\includegraphics[width=1.0\textwidth]{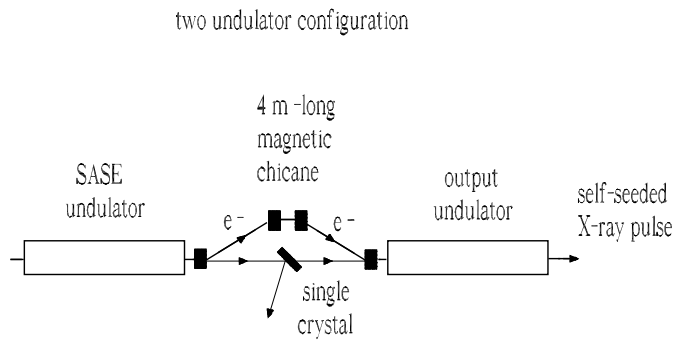}
\caption{Schematics of a single-crystal self-seeding scheme for hard
X-rays. It will rely on a diamond crystal,  C(400) reflection, in
Bragg geometry.} \label{SASE2_1}
\end{figure}

\begin{figure}[tb]
\includegraphics[width=1.0\textwidth]{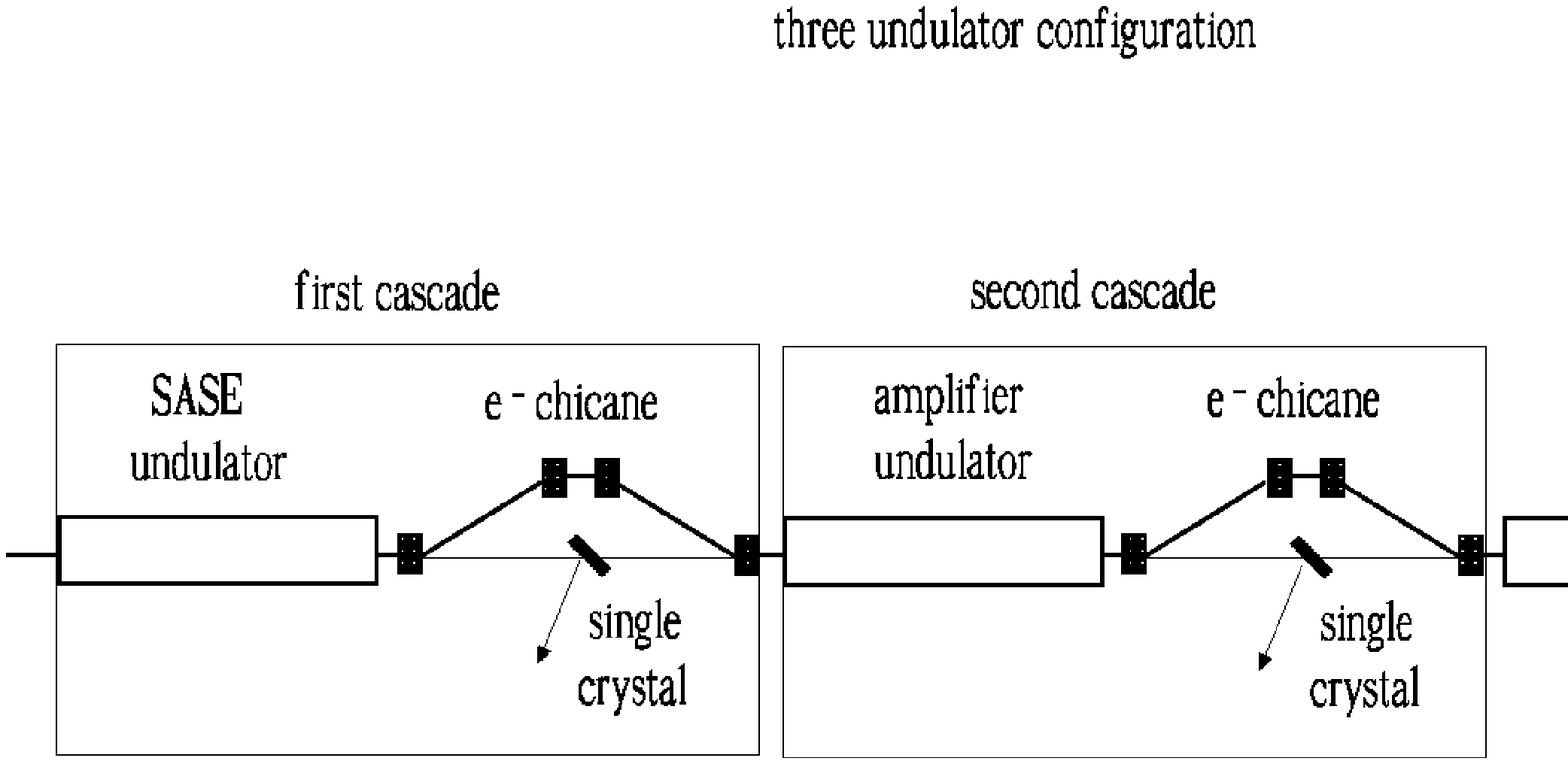}
\caption{Schematics of a two-cascade self-seeding scheme with single
crystal monochromators. This amplification-monochromatization
cascade scheme is distinguished, in performance, by a high spectral
purity of the output radiation.} \label{SASE2_2}
\end{figure}

The self-seeding scheme is a promising approach to significantly
narrow the SASE bandwidth and to produce nearly transform-limited
pulses \cite{SELF}-\cite{WUFEL2}. As shown in Fig. \ref{SASE2_1},
the self-seeding setup consists of two undulators separated by a
photon monochromator and an electron bypass beam line, normally a
$4$-dipole chicane. The two undulators are resonant to the same
radiation wavelength. The SASE radiation generated by the first
undulator passes through the narrow-band monochromator to create  a
transform-limited pulse, which is then used as a coherent seed in
the second undulator. Chromatic dispersion effects in the bypass
chicane smear out the microbunching in the electron bunch produces
by the SASE lasing in the first undulator. The electrons and the
monochromatized photon beam are recombined at the entrance of the
second undulator, and the radiation is amplified by the electron
bunch in the second undulator, until saturation is reached. The
required seed power at the beginning of the second undulator must
dominate over the shot noise power within the gain bandpass, which
is order of a few kW.

For hard X-ray self-seeding, a monochromator is normally configured
with crystals in the Bragg geometry. A conventional $4$-crystal
monochromator introduces an optical delay at least a few
millimiters, which has to be compensated with the introduction of a
long electron bypass. To avoid such a long chicane, a simpler
self-seeding scheme was proposed in \cite{OURX}-\cite{OURY6}, which
uses the transmitted X-ray beam from a single crystal to seed the
same bunch (see Fig. \ref{SASE2_1}).

With radiation beam monochromatized down to the Fourier transform
limit, a variety of very different techniques leading to further
improvement of the X-ray FEL performance become feasible. Despite
the unprecedented increase in peak power of the X-ray pulses for
SASE X-ray FELs, some applications, including single biomolecule
imaging,  may require still higher photon flux. The most promising
way to extract more FEL power than that at saturation is by tapering
the magnetic field of the undulator \cite{TAP1}-\cite{TAP4}. Also, a
significant increase in power is achievable by starting the FEL
process from monochromatic seed rather than from noise
\cite{FAWL}-\cite{WANG}. Recently we proposed to make use of a
single crystal self-seeding scheme to be installed in the
tunable-gap baseline undulator at the European XFEL to create a
source capable of delivering transform-limited X-ray pulses at
extraordinary peak power level ($0.4$ TW) \cite{OURY3}. This single
crystal self-seeding scheme was examined in \cite{OURY3} for the
European XFEL by using ideal electron beam characteristics.

A typical X-ray FEL beam after acceleration and compression is
usually not as simple as the case discussed in that paper, which
deals with a Gaussian beam distribution and ignores wakefield
effects. In the present paper we extend our consideration to a more
realistic electron beam distribution at the undulator entrance. In
particular, we propose a study of the performance of single-crystal
self-seeding scheme for the European XFEL, based on start-to-end
simulations, and accounting for undulator wakefields \cite{S2ER}. We
optimize our self-seeding setup, based on the results of
start-to-end simulations for an electron beam with $30$ pC charge.
Our analysis indicates that the self-seeding scheme is not
significantly affected by non-ideal electron phase space
distribution, and yields about the same performance as in the case
for an electron beam with ideal properties. Simulations show that
the FEL power of the transform-limited X-ray pulses may be increased
up to $0.4$ TW by operating with a tapered baseline undulator. In
particular, it is possible to create a source capable of delivering
fully-coherent, $7$ fs (FWHM) X-ray  pulses with $2 \cdot 10^{12}$
photons per pulse at a wavelength of $0.15$ nm.

\section{Possible single-crystal self-seeding scheme for the European XFEL
baseline undulator}

In its simplest configuration, a self-seeding FEL consists of an
input and output undulator separated by a monochromator. With
reference Fig. \ref{SASE2_2} we discuss the simplest two-undulator
configuration case for the single-crystal self-seeding setup. The
first undulator operates in the linear high-gain regime starting
from the shot-noise in the electron beam. After the first undulator
the output SASE radiation passes through the monochromator, which
reduces the bandwidth. A distinguishing feature of our proposed
setup is that it takes advantage of a single crystal in Bragg
transmission geometry, instead of a fixed exit four-crystal
monochromator. Due to nearly $100 \%$ in-band reflectivity, in the
frequency domain the crystal works as a notch filter for the
transmitted X-rays and generates a monochromatized wake in the time
domain. The delay time of this monochromatized pulse from the main
FEL pulse is determined by the width of the notch filter. The peak
of the wake is about $10$ MW and is delayed from the main SASE pulse
by $6-7~\mu$m. A weak magnetic chicane (about $5$ m long) is
sufficient to delay the electron bunch and to smear out the SASE
microbunching induced in the first part of the undulator. Since the
rms electron bunch length is about $1 ~\mu$m, after the chicane the
wake can be superimposed to the entire electron bunch, and acts as
effective seed in the second undulator. In order for the seed to
dominate the shot-noise generated in the second undulator, the SASE
FEL in the first undulator has to provide sufficiently high FEL
power to compensate for the power reduction associated with the
passage through the single-crystal monochromator.

In some experimental situations, the simplest two-undulator
configuration is not optimal. For example, the European XFELis
characterized by a very high repetition rate, and a peculiar bunch
structure, leading to important heat-loading of the monochromator
and limiting the maximal seed power from the first undulator part. A
possible extension of the two-undulator configuration consists in a
setup with three undulators separated by monochromators, Fig.
\ref{SASE2_2}. This amplification-monochromatization cascade scheme
is distinguished, in performance, by high spectral purity of the
output radiation  and a small heat-loading of the monochromator
crystals.

Finally, the most promising way to increase the output power of the
X-ray FEL is by tapering the magnetic field of the undulator.
Tapering consists in a slow reduction of the field strength of the
undulator in order to preserve the resonance wavelength, while the
kinetic energy of the electrons decreases due to the FEL process.
The undulator taper could be simply implemented as a step taper from
one undulator segment to the next. The magnetic field tapering is
provided by changing the undulator gap. A further increase in power
is achievable by starting the FEL process from the monochromatic
seed, rather than from noise. The reason is the higher degree of
coherence of the radiation in the seed case, thus involving, with
tapering, a larger portion of the bunch in the energy-wavelength
synchronism.

\begin{figure}[tb]
\includegraphics[width=1.0\textwidth]{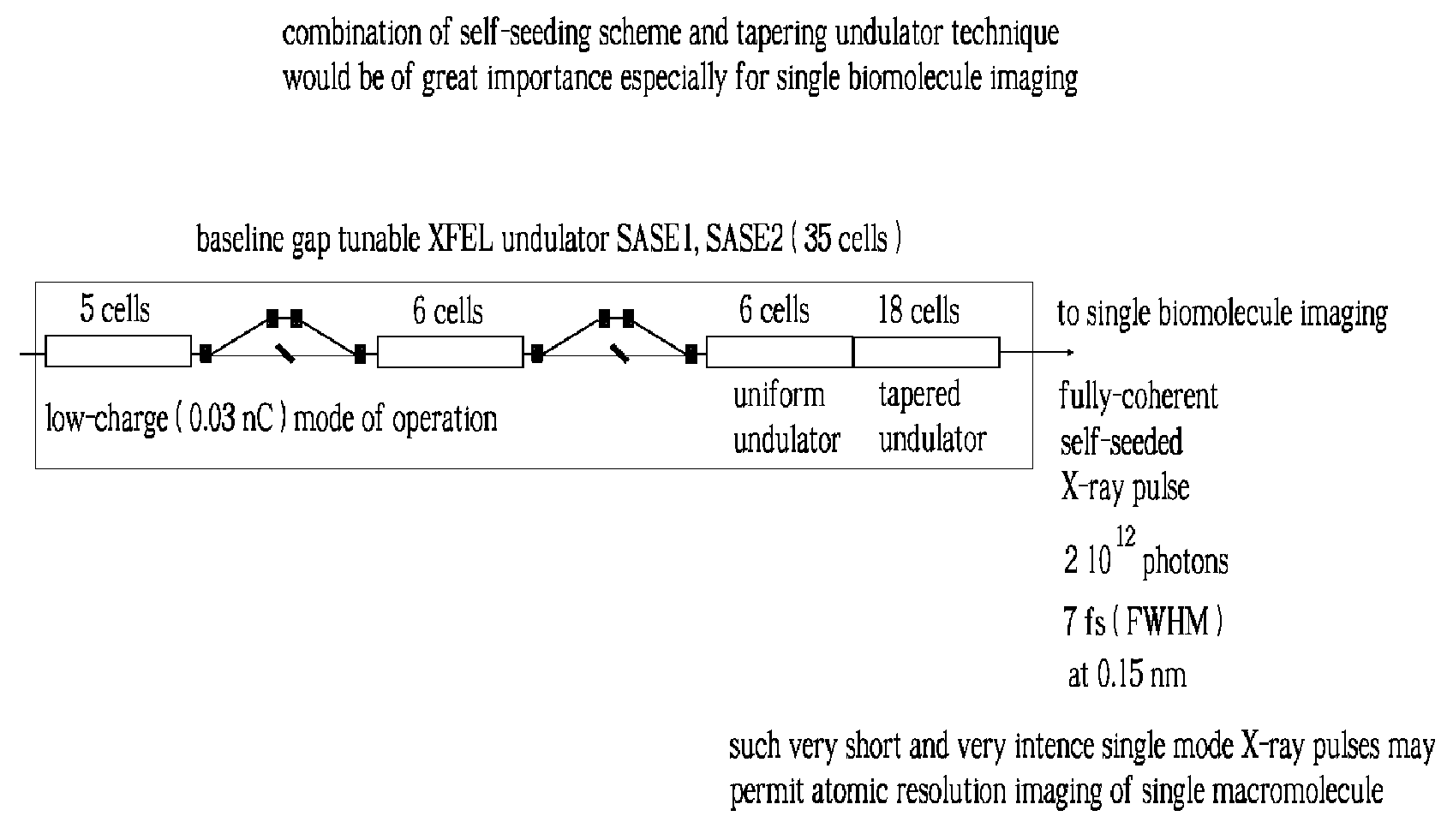}
\caption{Design of an undulator system for high power mode of
operation. The method exploits a combination of a cascade
self-seeding scheme with single crystal monochromators and an
undulator tapering technique. This scheme holds a great promise as a
source of X-ray radiation for applications such as single
biomolecule imaging.} \label{SASE2_3}
\end{figure}
Here we study a scheme for generating TW-level X-ray pulses in a
tapered undulator with the use of cascade self-seeding technique for
highly monochromatic seed generation. In this way the output power
of European X-ray FEL could be increased from baseline value of $20$
GW to about $400$ GW at $0.15$ nm wavelength. Fig. \ref{SASE2_3}
shows the design principle of our setup for high power mode of
operation. The scheme consists of two parts: a succession of two
amplification-monochromatization cascades and an output undulator in
which the monochromatic seed signal is amplified up to the TW
power-level.  Each cascade consists of an undulator, acting as an
amplifier, and a monochromator. Calculations show that in order not
to spoil the electron beam quality and to simultaneously solve the
heat-loading problem for the monochromator, the number of cells in
the first and second cascade should be $5$ and $6$ respectively. The
output undulator consists of two sections. The first section is
composed by an uniform undulator, the second section by a tapered
undulator. In the two cascades a nearly transform-limited FEL pulse
is produced, which is then  exponentially amplified passing through
the first uniform part of the output undulator. This section is long
enough ($6$ cells) to reach saturation, which yields about $20$ GW
power. Finally, in the second part of the output undulator the
monochromatic FEL output is enhanced up to $400$ GW with a $2 \%$
taper of the undulator parameter over the last 18 cells after
saturation.

For the European XFEL, our cascade scheme trivially satisfies heat
loading restrictions for the average power density, where the
situation is indeed much better than at third generation synchrotron
radiation sources. The energy per bunch impinging on the second
crystal, which bears the largest heat-load, can be estimated as $3
\mu$J (see section 3). One can easily estimate an average power of
$0.1$ W ($3$ $\mu$J times $2700$ pulses per train times $10$
trains/s). We consider a transverse rms dimension of the bunch of
about $15 ~\mu$m. Within a short ($5$ m -long) magnetic chicane, the
divergence of the X-ray radiation pulse is negligible, and the
radiation power is distributed as the electron bunch. As a result we
obtain illuminated crystal area of $1.2 \cdot 10^{-3}~
\mathrm{mm}^2$. This corresponds to a normal incident power density
of about $100 \mathrm{W}/\mathrm{mm}^2$ at the position of the
second monochromator, about an order of magnitude smaller compared
to the average power density at monochromators of third generation
synchrotron sources.

However, the European XFEL differs compared to third generation
sources in the very specific time diagram, which foresees the
production of about $10$ trains of pulses per second, each train
consisting of $2700$ pulses. In this case, the average power density
along a single pulse train is the meaningful figure of merit, rather
than the above-mentioned average power density.  The average power
within a single bunch train can be estimated by multiplying the
energy by about $3000$ pulses composing a single train and dividing
by a temporal duration of a train, which is $0.6$ ms. One obtains
power of $20$ W. Considering illuminated area as $10^{-3}
\mathrm{mm}^2$, as before, we obtain a power density of about $10
\mathrm{kW}/\mathrm{mm}^2$ within a single train, at normal
incidence. Such heat-load is an order of magnitude smaller than what
is foreseen at monochromators for the SASE2 baseline, where a
diamond crystal with the same thickness ($0.1$ mm) is planned to be
used.

\section{Simulations}

\subsection{Start-to-end electron beam simulations}

In this Section we report on a feasibility study performed with the
help of the FEL code GENESIS 1.3 \cite{GENE} running on a parallel
machine. We will present a feasibility study for a short-pulse mode
of operation of the SASE1 and SASE2 FEL lines of the European XFEL,
based on a statistical analysis consisting of $100$ runs. The
overall beam parameters used in the simulations are presented in
Table \ref{tt1}. We refer to the setup in Fig. \ref{SASE2_3}.

\begin{table}
\caption{Parameters for the low-charge mode of operation at the
European XFEL used in this paper.}

\begin{small}\begin{tabular}{ l c c}
\hline & ~ Units &  ~ \\ \hline
Undulator period      & mm                  & 40     \\
Periods per cell      & -                   & 125   \\
K parameter (rms)     & -                   & 2.15  \\
Total number of cells & -                   & 35    \\
Intersection length   & m                   & 1.1   \\
Wavelength            & nm                  & 0.15  \\
Energy                & GeV                 & 14.0 \\
Charge                & pC                  & 28\\
\hline
\end{tabular}\end{small}
\label{tt1}
\end{table}

\begin{figure}[tb]
\includegraphics[width=0.5\textwidth]{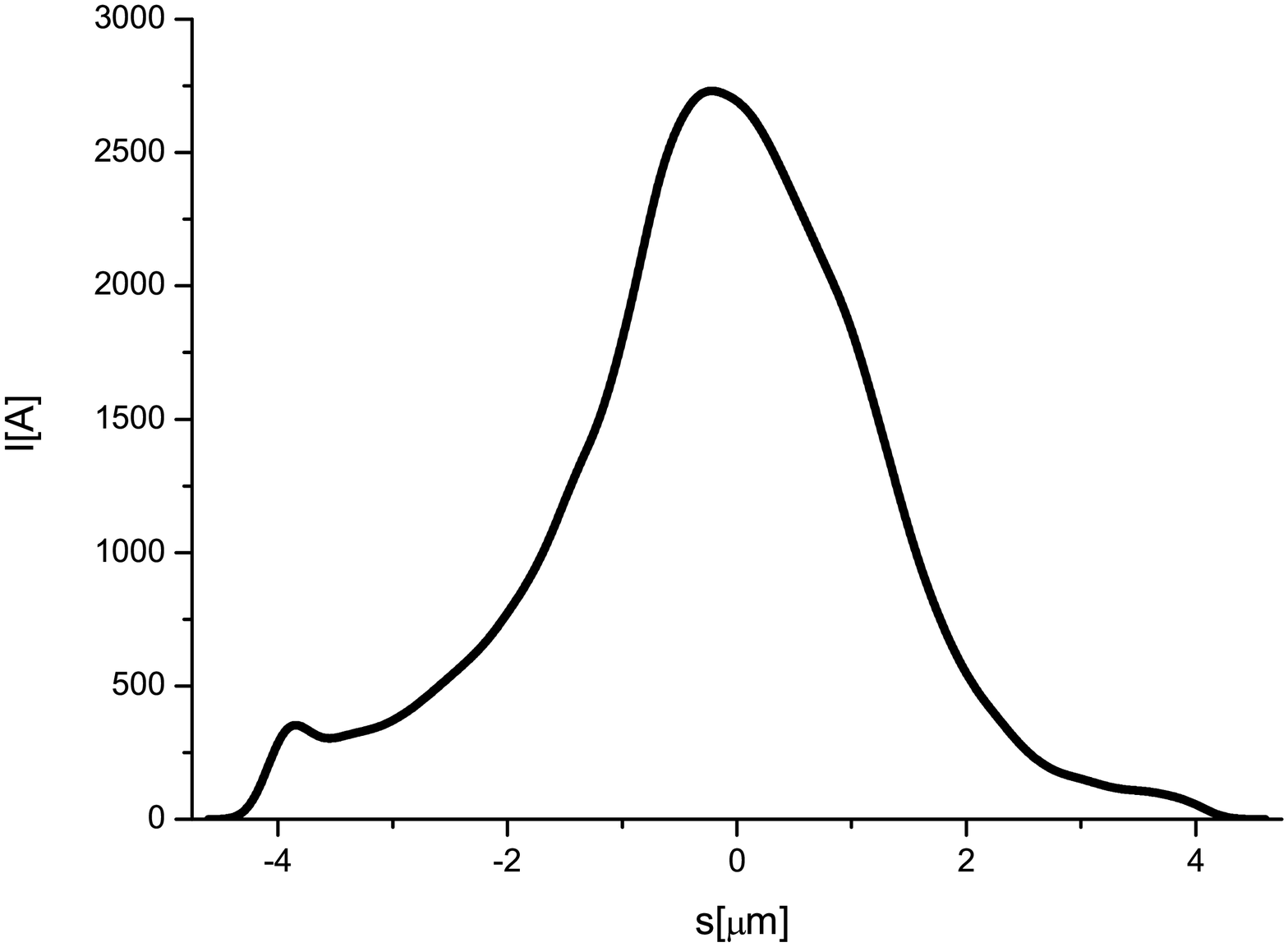}
\includegraphics[width=0.5\textwidth]{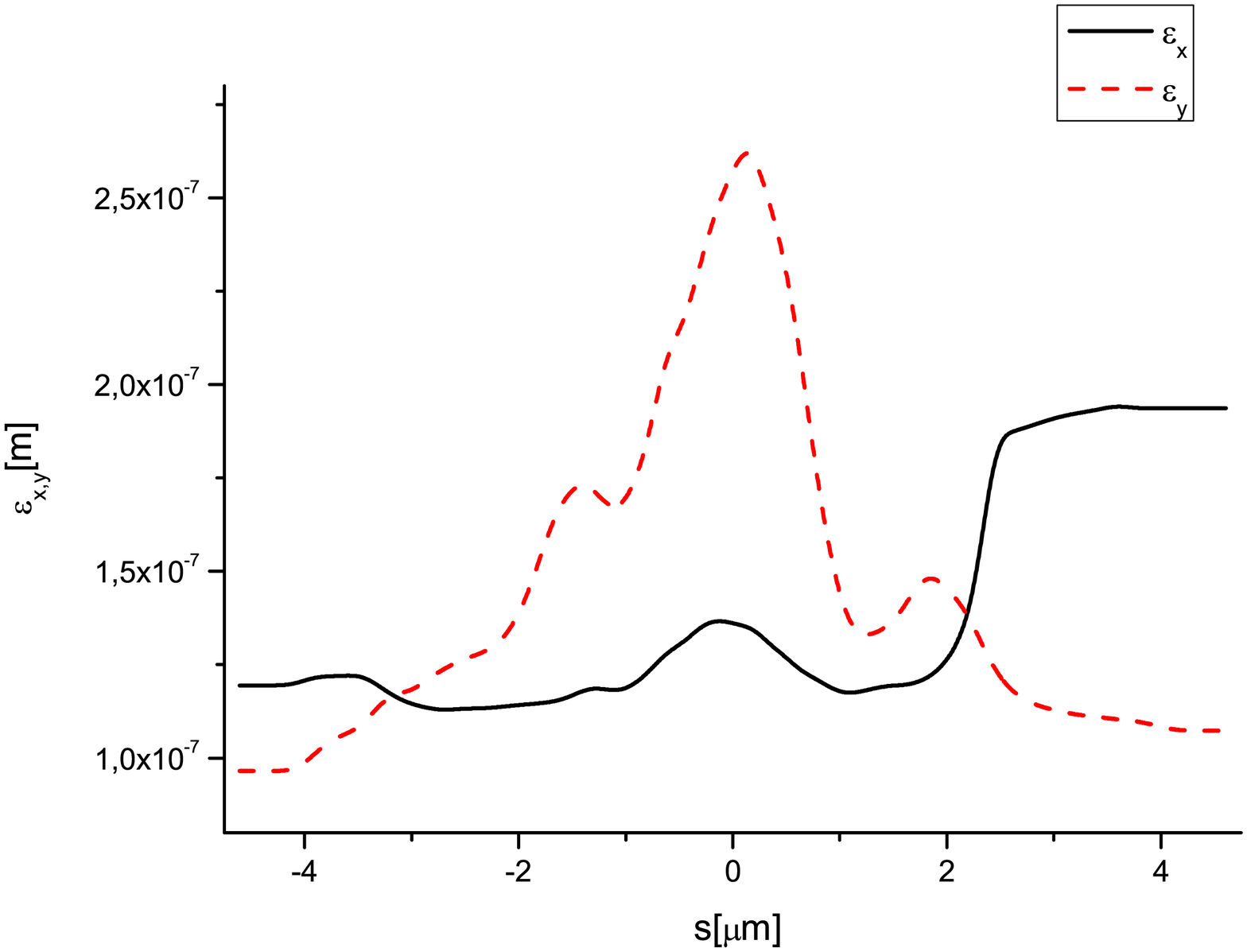}
\includegraphics[width=0.5\textwidth]{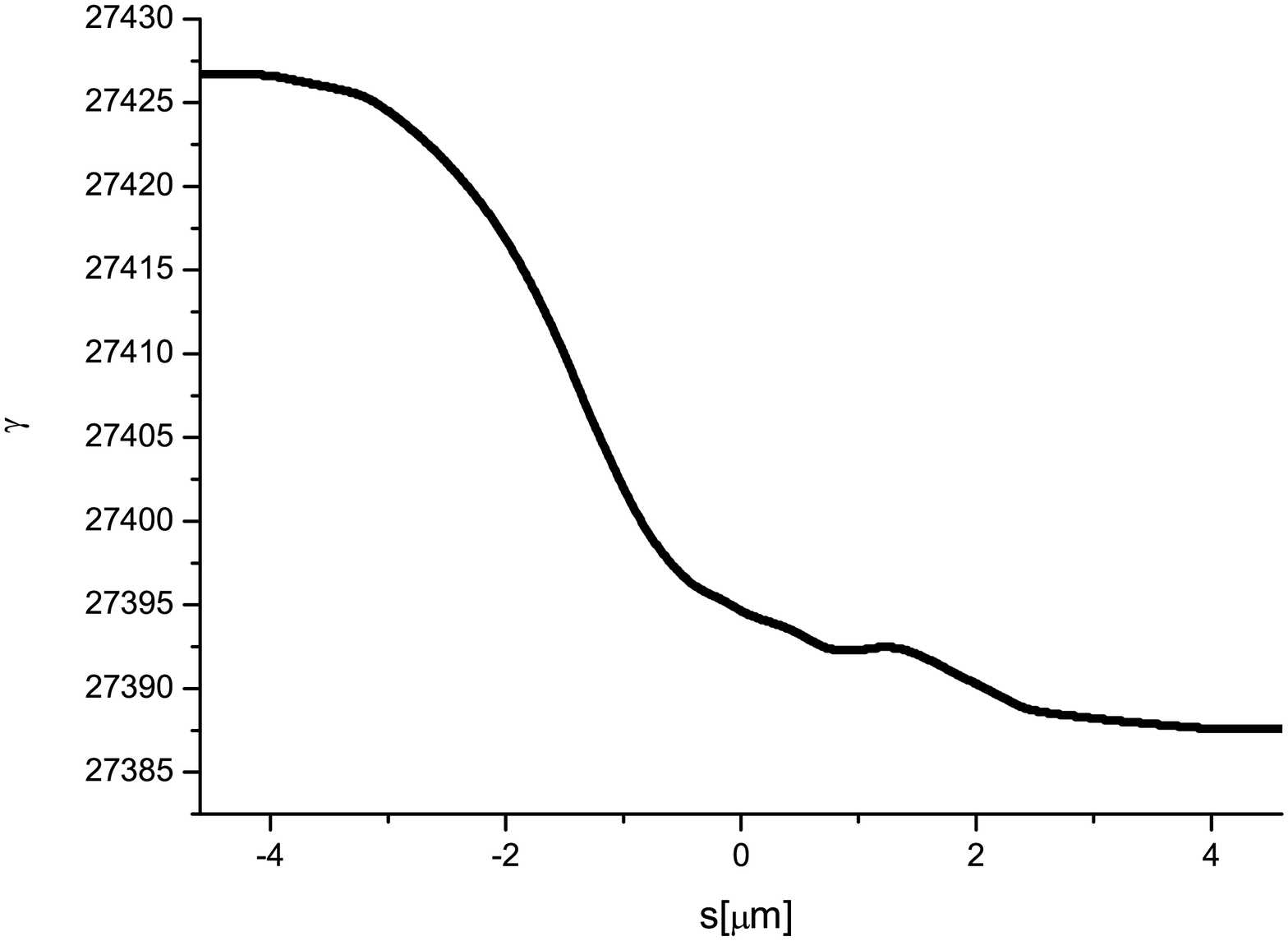}
\includegraphics[width=0.5\textwidth]{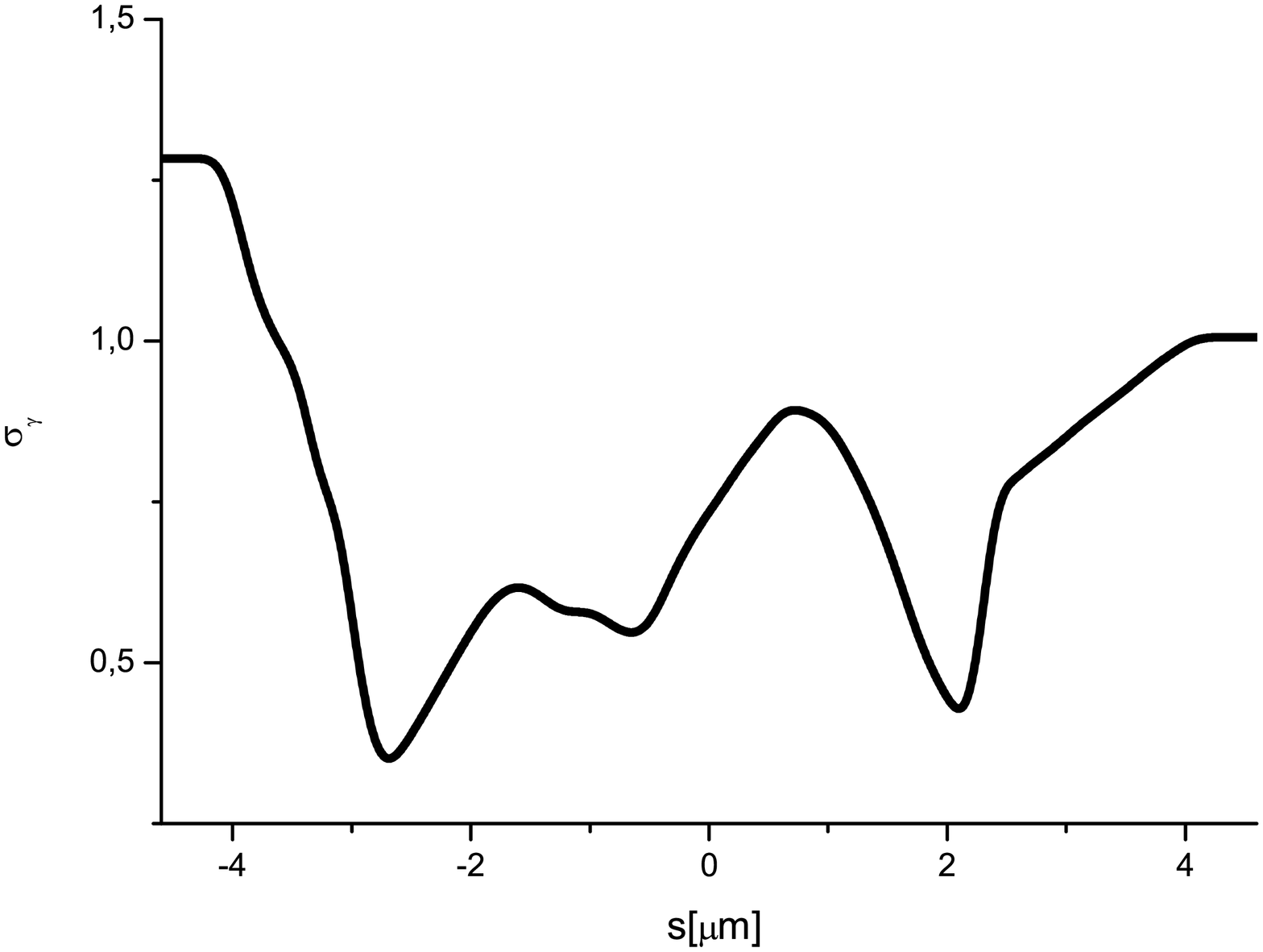}
\caption{Results from electron beam start-to-end simulations at the
entrance of SASE1 and SASE2 \cite{S2ER}. (Top Left) Current profile.
(Top Right) Normalized emittance as a function of the position
inside the electron beam. (Bottom Left) Energy profile along the
beam. (Bottom right) Electron beam energy spread profile.}
\label{s2E}
\end{figure}
\begin{figure}[tb]
\includegraphics[width=1.0\textwidth]{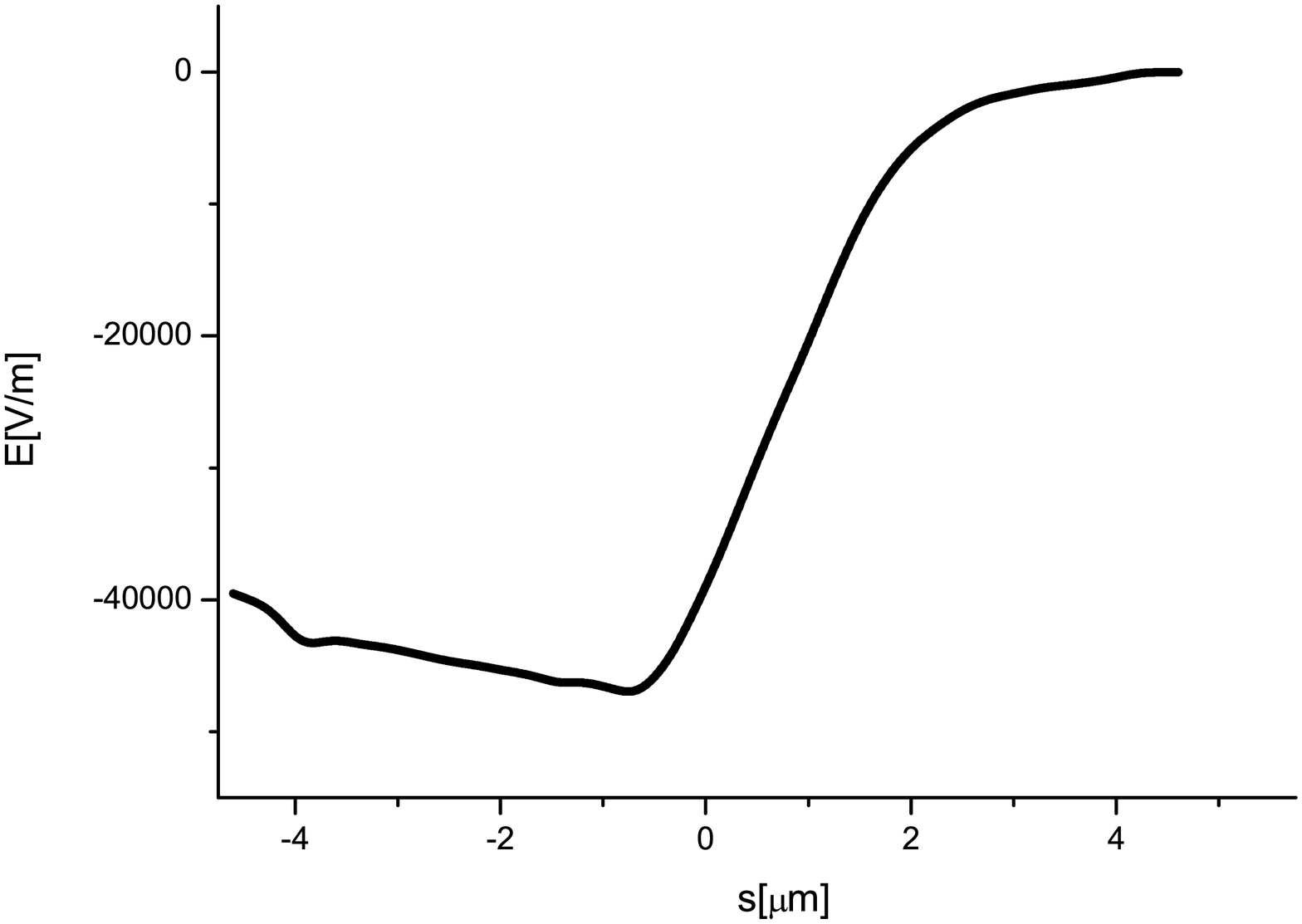}
\caption{Resistive wakefields in the SASE 1 (SASE 2) undulator
\cite{S2ER}} \label{wake}
\end{figure}

The expected beam parameters at the entrance of the SASE1 and SASE2
undulators are shown in Fig. \ref{s2E}, \cite{S2ER}. Wakes inside
the undulators are also accounted for and expected to obey the
dependence in Fig. \ref{wake}, \cite{S2ER}.

\begin{figure}[tb]
\includegraphics[width=0.5\textwidth]{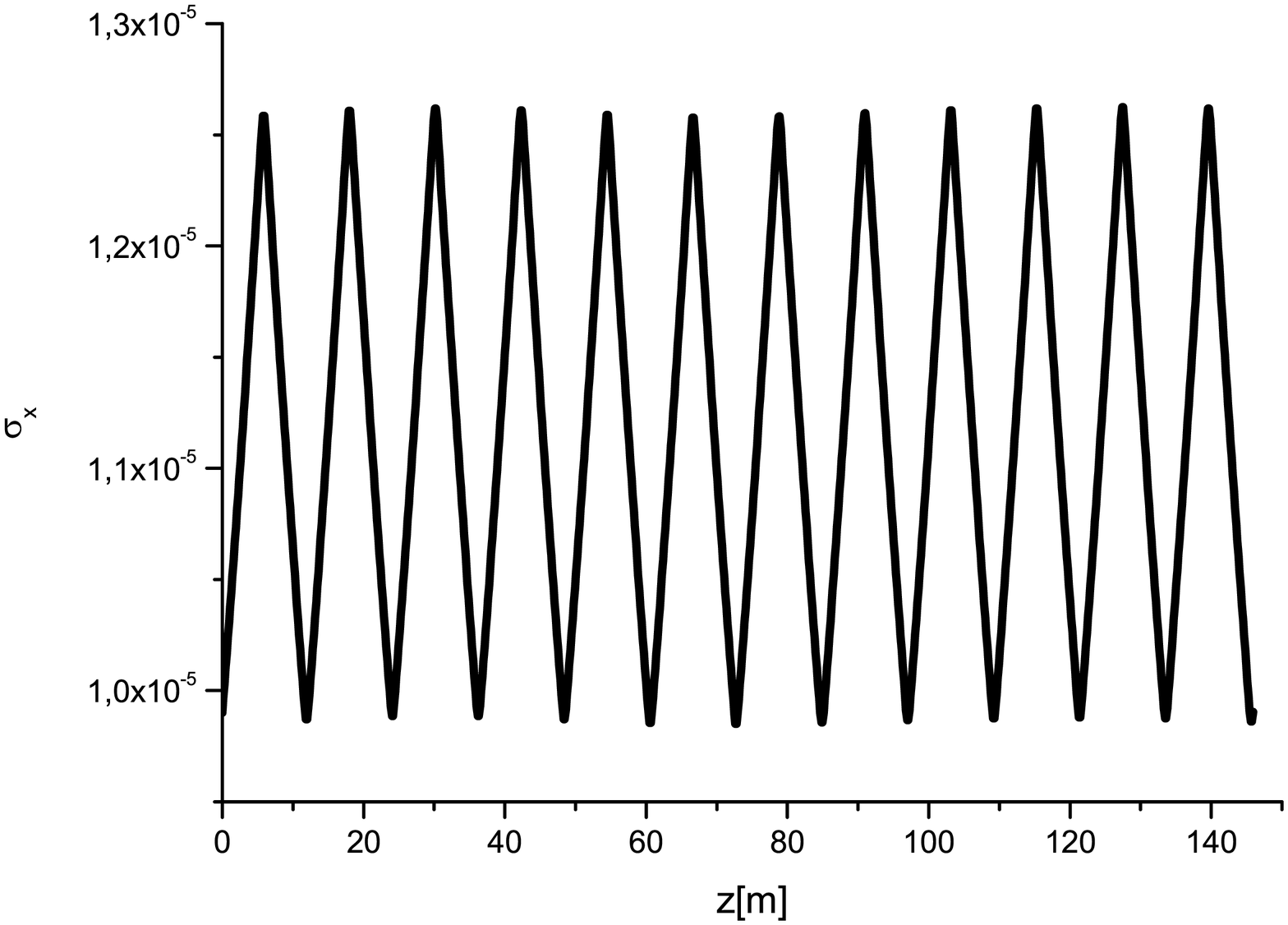}
\includegraphics[width=0.5\textwidth]{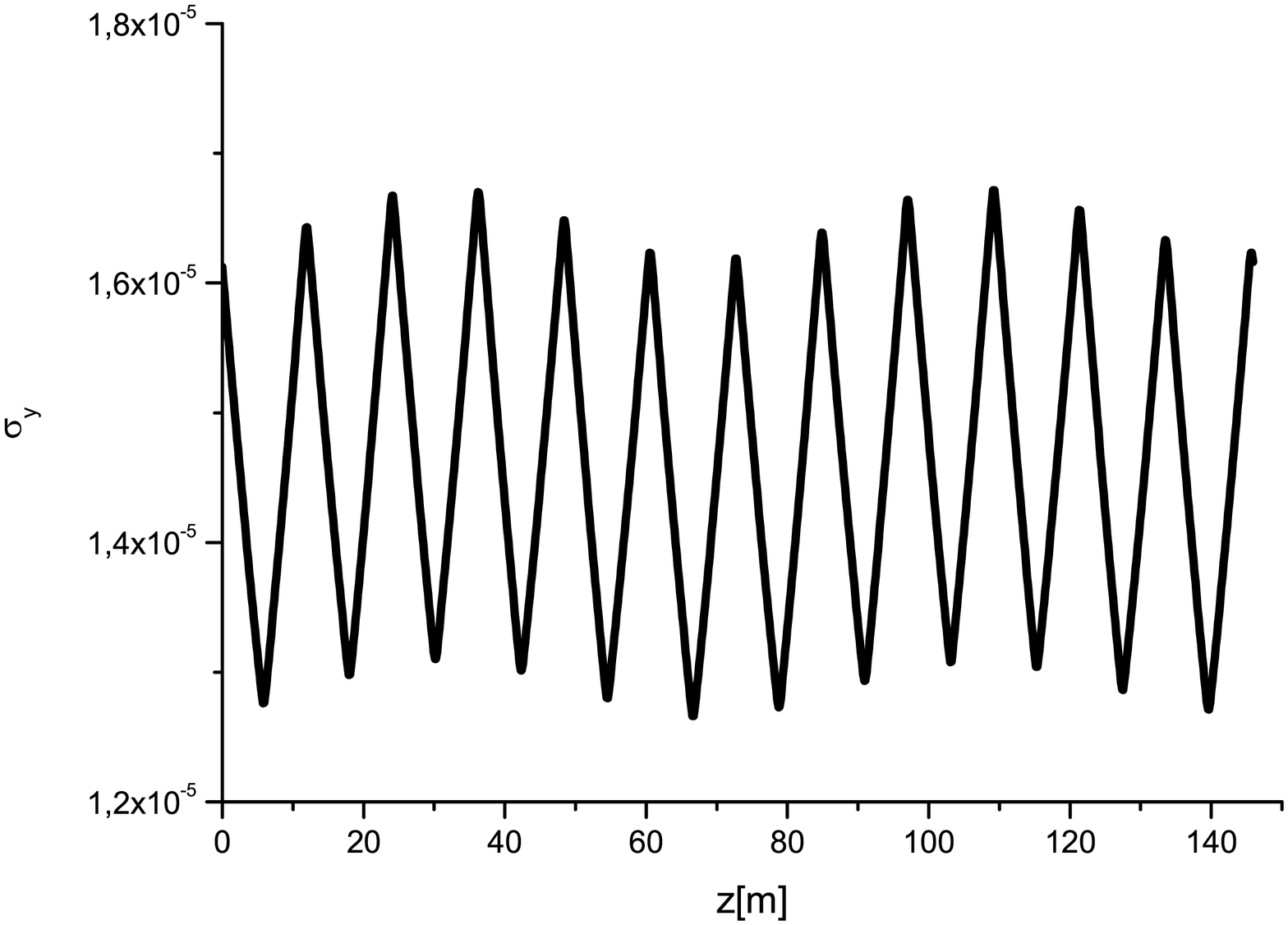}
\caption{Evolution of the rms horizontal (left plot) and vertical
(right plot) beam size as a function of the distance along the setup
calculated through GENESIS. These plots refer to the position within
the bunch where the current is maximal. The normalized horizontal
emittance is $\varepsilon_x = 1.37\cdot 10^{-7}$ m. The normalized
vertical emittance is $\varepsilon_y = 2.4\cdot 10^{-7}$ m.}
\label{sigperp}
\end{figure}
The evolution of the rms horizontal and vertical beamsize as a
function of the distance along the setup can be calculated through
Genesis, and is shown in Fig. \ref{sigperp}. The figure shows the
evolution for the position of maximal current in the bunch, where
normalized horizontal and vertical emittances are, respectively,
$\varepsilon_x = 1.37\cdot 10^{-7}$ m and $\varepsilon_y = 2.4\cdot
10^{-7}$ m. Inspection of Fig. \ref{sigperp} shows a little mismatch
in the vertical direction $y$, which is not present in the
horizontal direction, and is due to the fact that undulator focusing
forces are in the vertical direction only. The mismatch is
automatically accounted for in the Genesis simulations. Since it
turns out to be not significant, we keep a uniform FODO lattice
focusing system.

\subsection{SASE radiator and first crystal monochromator}

\begin{figure}[tb]
\includegraphics[width=0.5\textwidth]{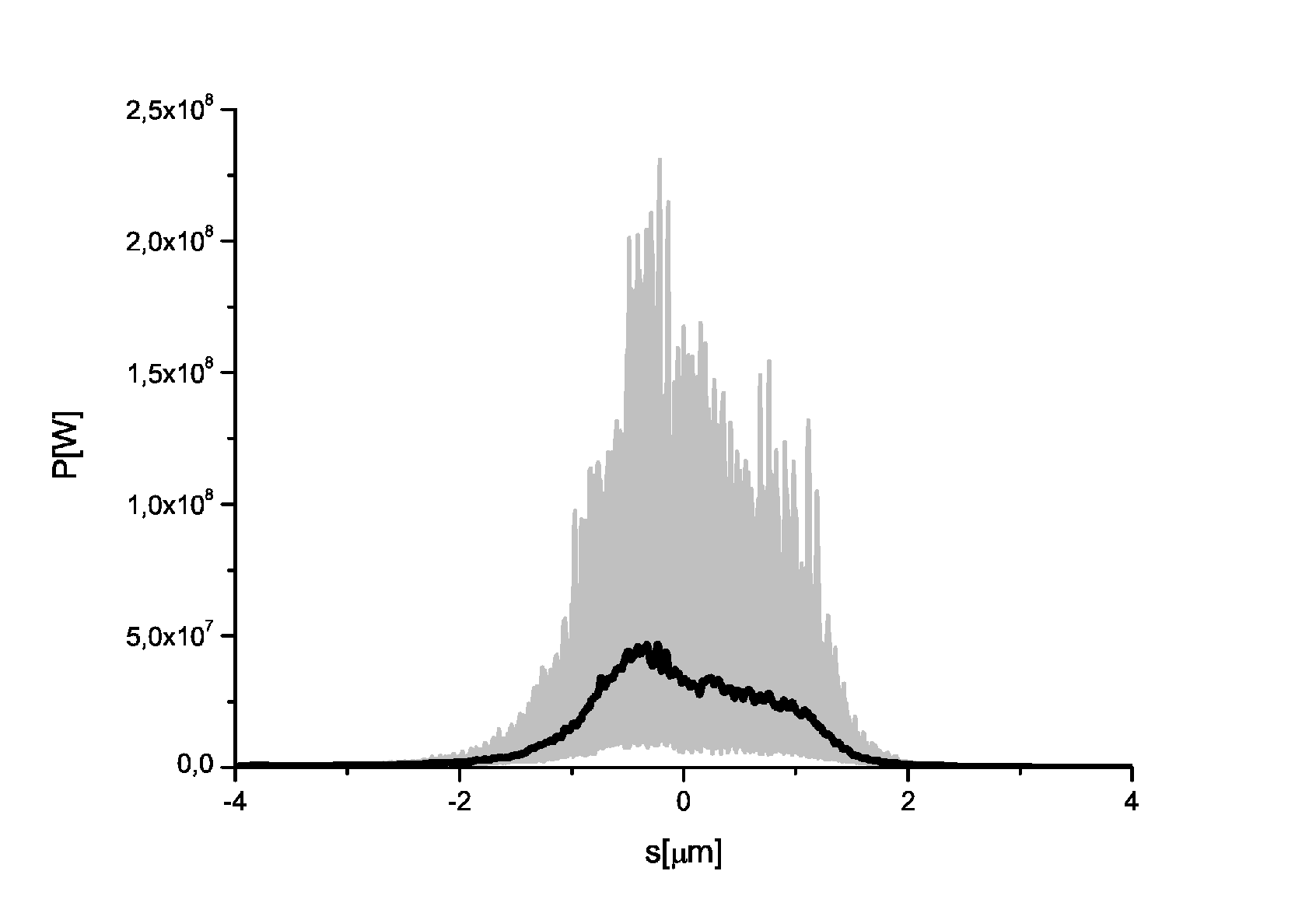}
\includegraphics[width=0.5\textwidth]{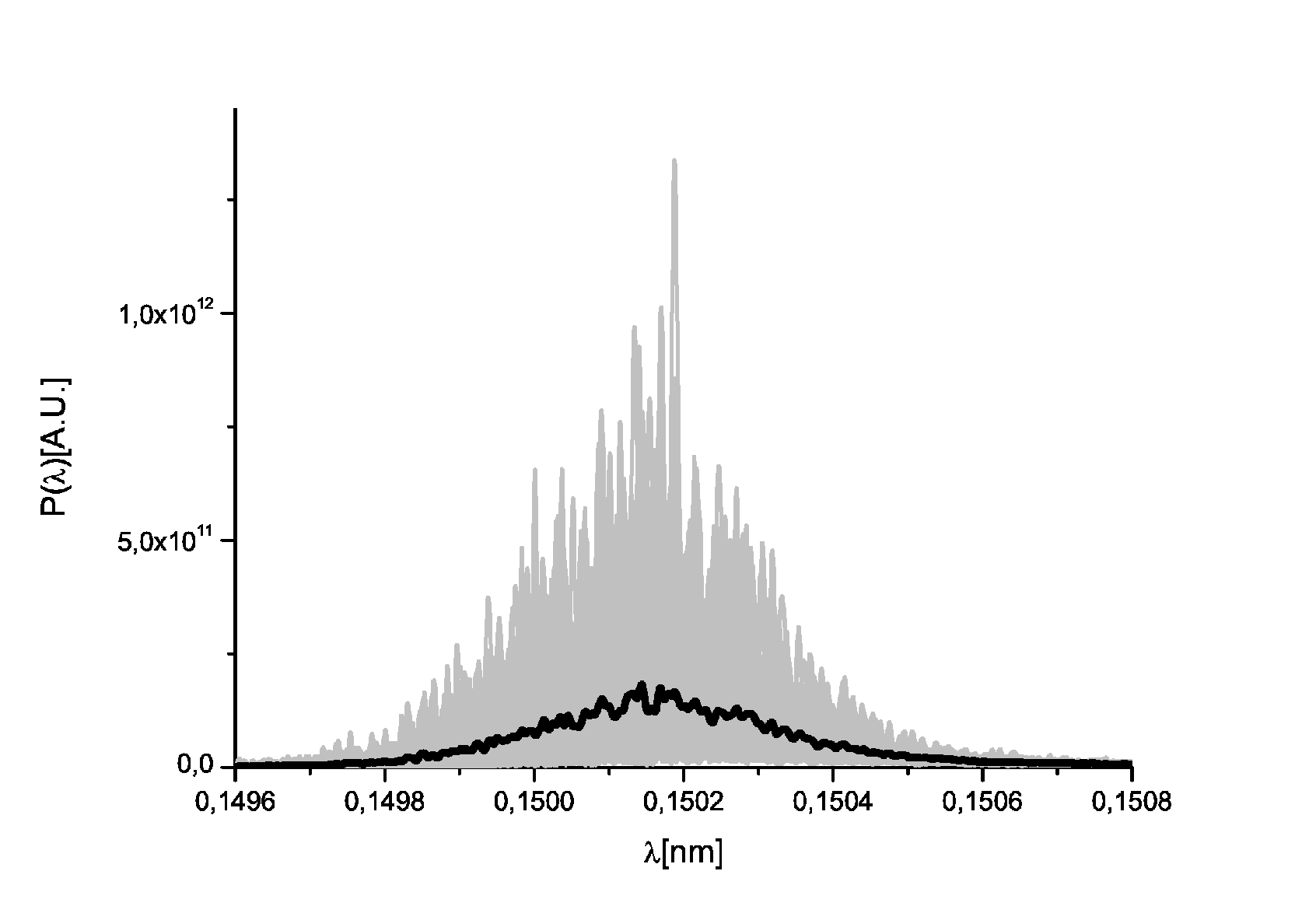}
\caption{(Left plot) SASE input power before the first crystal.
(Right plot) SASE input spectrum before the first crystal. Grey
lines refer to single shot realizations, the black line refers to
the average over a hundred realizations.} \label{seedin1}
\end{figure}
According to the scheme in Fig. \ref{SASE2_3}, the electron beam
first goes through 5 undulator cells radiating in the SASE mode. The
power and spectrum from the SASE radiator are shown in Fig.
\ref{seedin1}.

\begin{figure}[tb]
\includegraphics[width=0.5\textwidth]{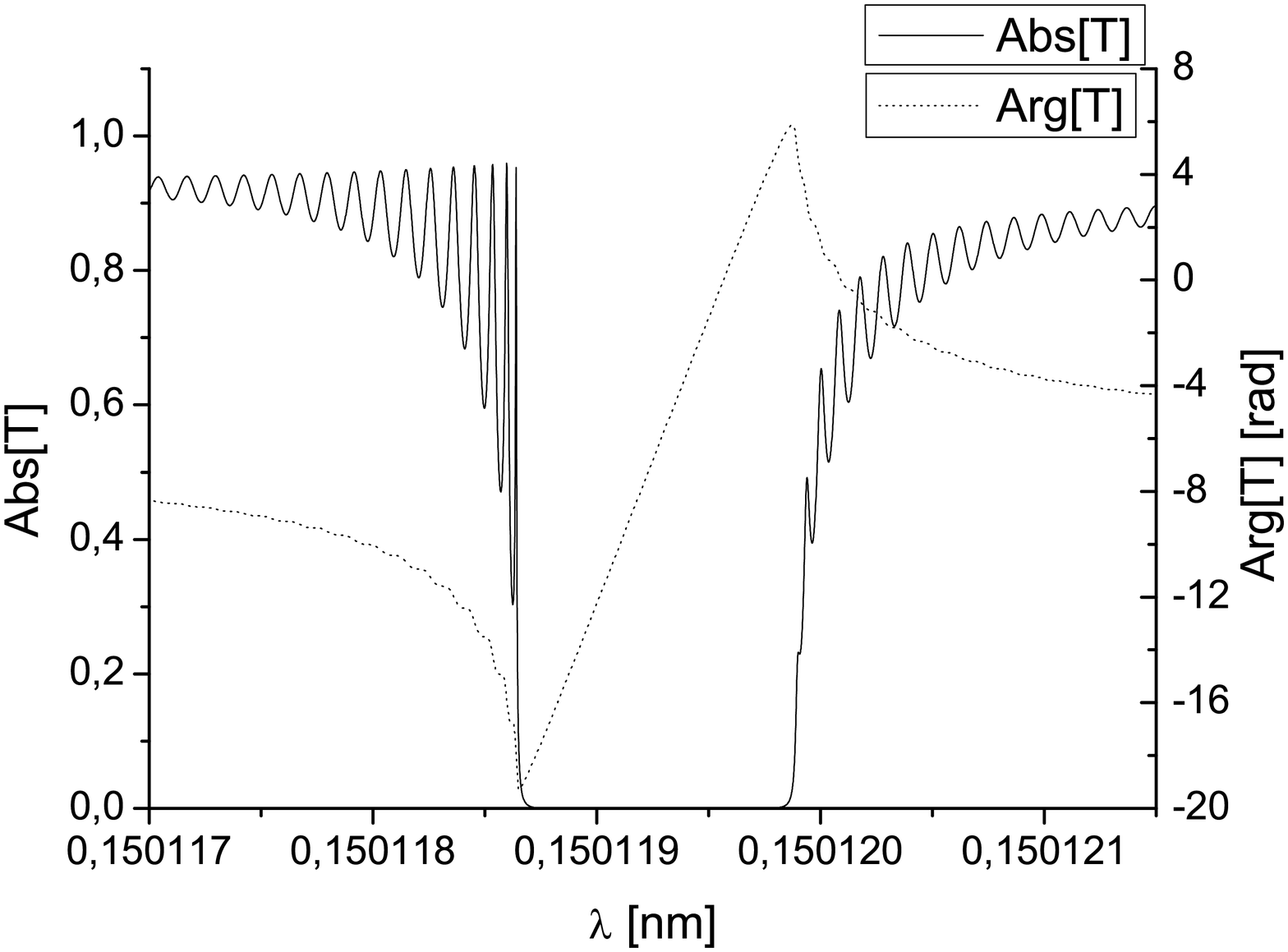}
\includegraphics[width=0.5\textwidth]{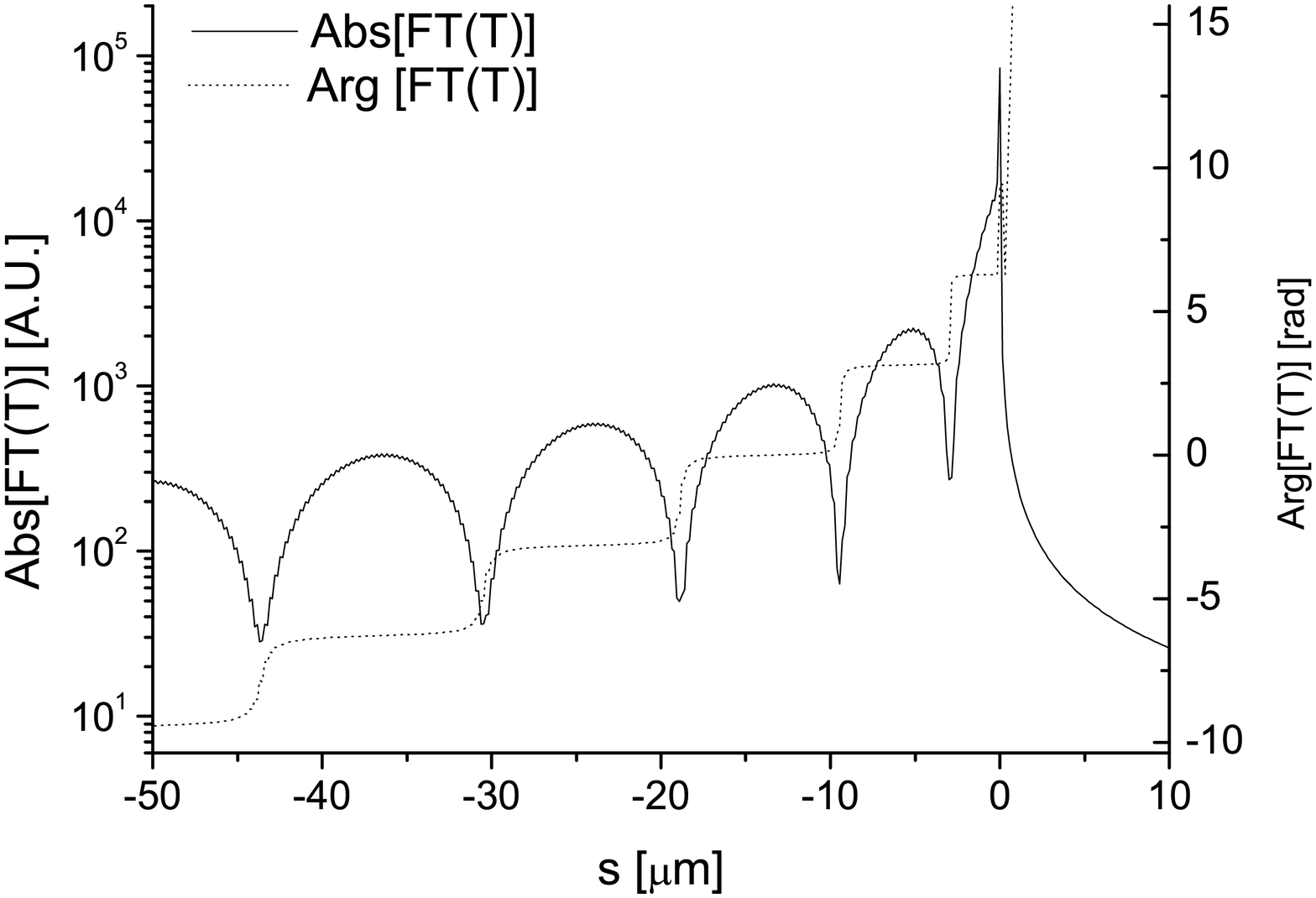}
\caption{(Left plot) Transmissivity (sigma polarization) relevant to
the Bragg 400 diffraction of X-rays at $0.15$ nm from a Diamond
crystal with a thickness of $0.1$mm. (Right plot) Fourier Transform
of the transmissivity in the left plot.} \label{TTT}
\end{figure}
\begin{figure}[tb]
\includegraphics[width=0.5\textwidth]{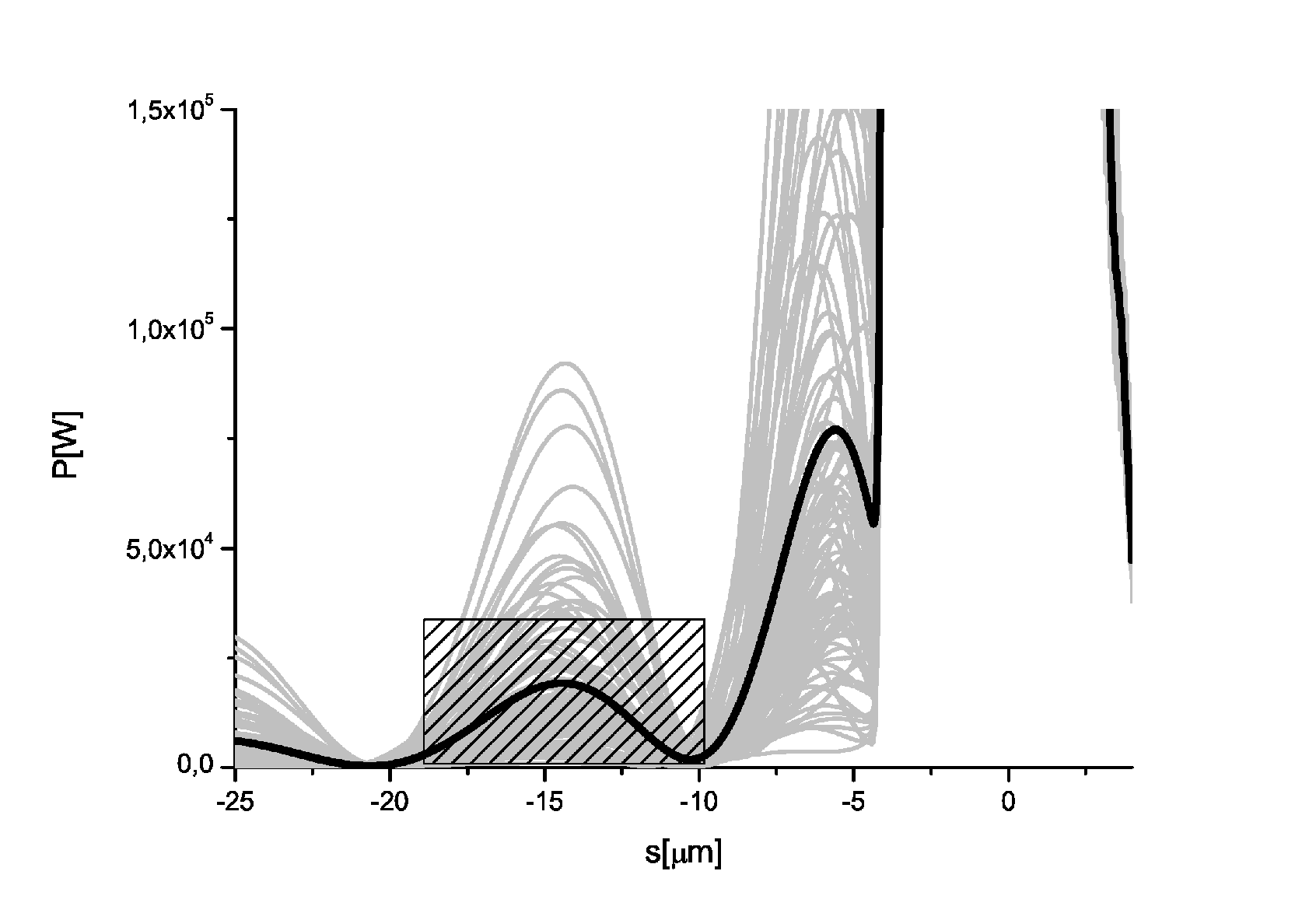}
\includegraphics[width=0.5\textwidth]{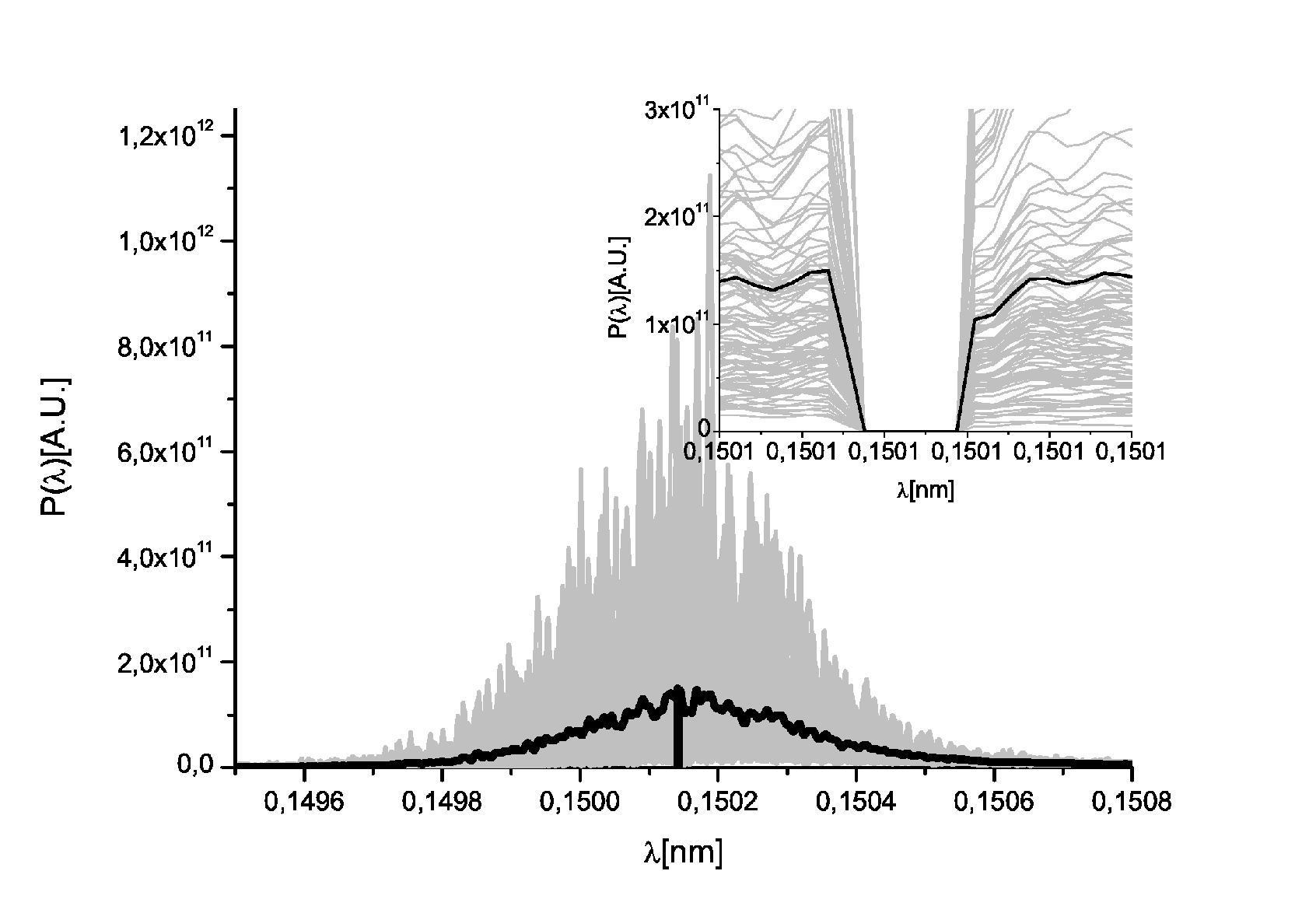}
\caption{(Left plot) Seed power after the first crystal. (Right
plot) The effect of the first crystal on the SASE spectrum. Grey
lines refer to single shot realizations, the black line refers to
the average over a hundred realizations.} \label{Seed1}
\end{figure}
After the first undulator, the output SASE radiation passes through
the monochromator, consisting of a crystals in the Bragg
transmission geometry, Fig. \ref{SASE2_2}. The crystal operates as
bandstop filters for the transmitted X-ray SASE radiation pulse,
Fig. \ref{TTT}, left plot. The modulus and the phase of the
transmissivity is calculated with the help of the dynamical theory
of X-ray diffraction. The output spectrum is given by the product of
$|T|^2$ with the initial SASE spectrum in Fig. \ref{seedin1}, and is
shown in Fig. \ref{Seed1}, right plot. The temporal waveform of the
transmitted radiation pulse shows a long tail, shown in Fig.
\ref{Seed1}, right plot, which can be used for seeding the electron
bunch after the chicane. The presence of such tail can be seen as a
direct consequence of the Fourier transform relations. In
particular, the Fourier transform of the transmissivity is shown in
Fig. \ref{TTT}, right plot. The modulus of the Fourier transform of
$T$ exhibits oscillations which are inversely proportional to the
bandwidth of the absorption line in the transmittance spectrum. This
is easy to explain in terms of the Fourier transform of a square
function. The sharp peak at $s=0$ corresponds to the long, nearly
constant ends of $|T|$ at large and small frequencies, which can be
seen in Fig. \ref{TTT}, left plot. Finally, the presence of a
particular phase in $T$ is related with the fulfillment of the
causality requirement, so that the modulus of the Fourier transform
of $T$ vanishes at positive values of $s$.

\begin{figure}[tb]
\includegraphics[width=1.0\textwidth]{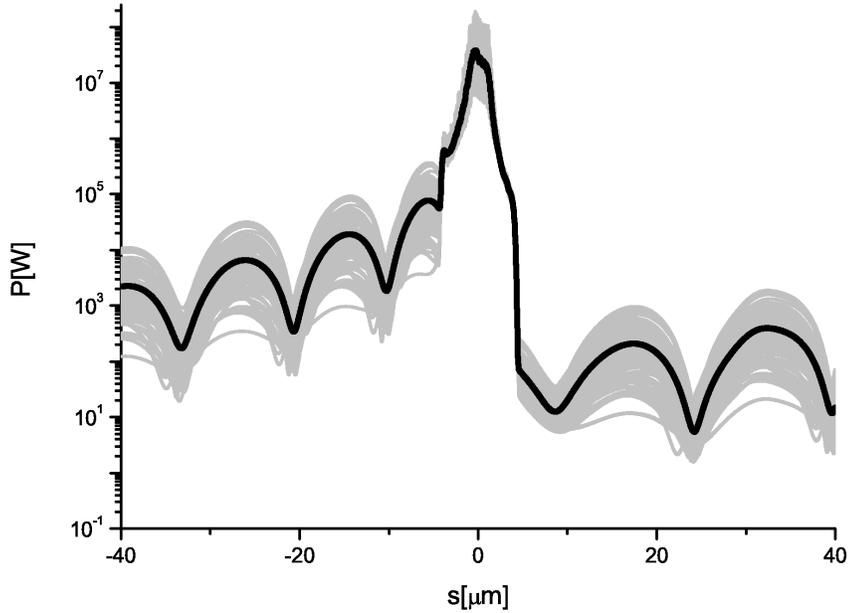}
\caption{Logarithmic plot of the seed power after the first crystal.
Grey lines refer to single shot realizations, the black line refers
to the average over a hundred realizations.} \label{Logseed1}
\end{figure}
The temporal wavefront of the transmitted radiation is obtained
convolving Fig. \ref{TTT}, right plot, with the SASE field in the
time domain. In other words, the Fourier transform of the
transmissivity can be seen as the input response of a filter.
Since we consider the low charge mode of operation, the oscillations
of the modulus of the Fourier transform of $T$ extend for a length
which is greater than the bunch length, so that the length of the
oscillations in Fig. \ref{Seed1}, right plot, are nearly inversely
proportional to the bandwidth of the absorption line in the
transmittance spectrum, as those in Fig. \ref{TTT}, left plot, as it
can be seen by comparing that plot with Fig. \ref{Logseed1}. Very
small but visible departure from the zero power level (the
theoretical value due to causality) appears in the right-hand side
of Fig. \ref{Logseed1}, for $s>0$. This accuracy is acceptable for
most purposes.

\begin{figure}[tb]
\includegraphics[width=1.0\textwidth]{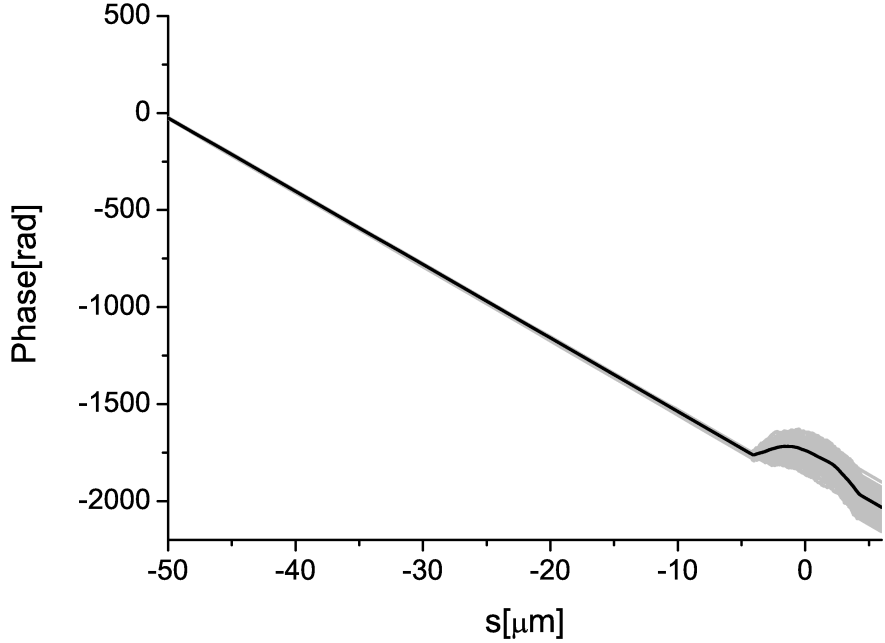}
\caption{Phase relation along the FEL pulse after the first crystal.
The large linear chirp is due to the definition of the reference
carrier frequency by Genesis, and is artificial. Grey lines refer to
single shot realizations, the black line refers to the average over
a hundred realizations.} \label{phaseT}
\end{figure}

\begin{figure}[tb]
\includegraphics[width=0.5\textwidth]{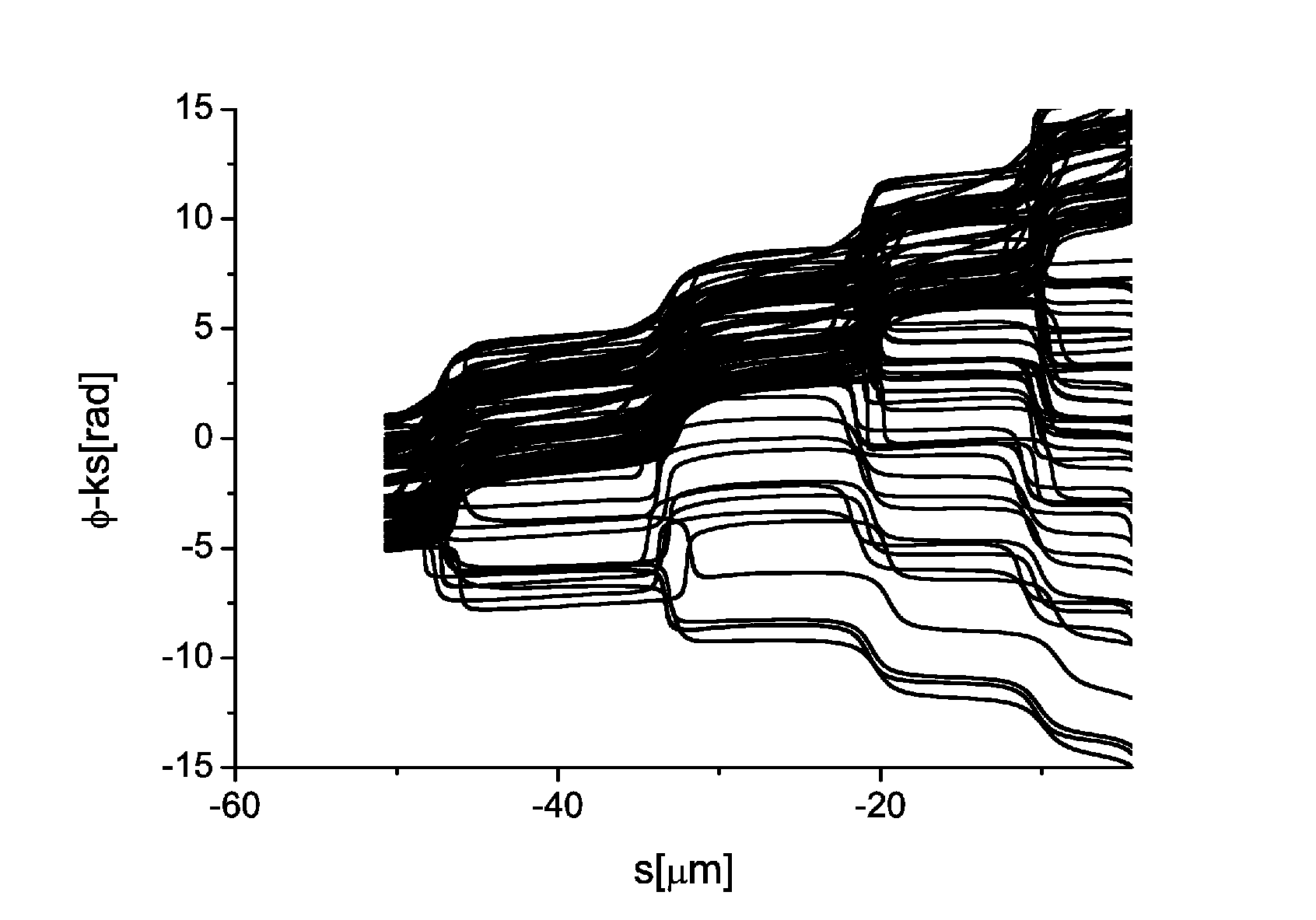}
\includegraphics[width=0.5\textwidth]{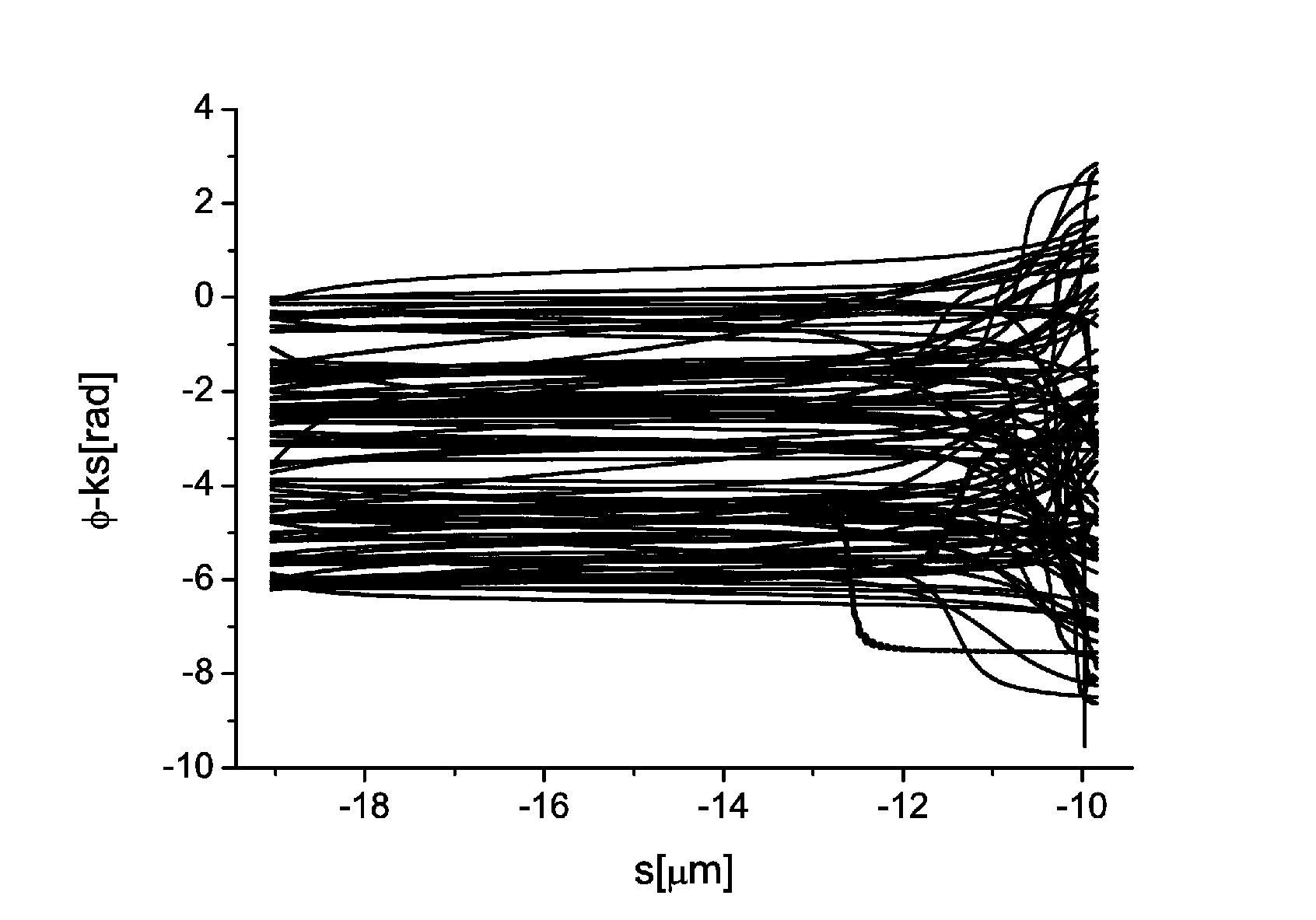}
\caption{(Left plot) Phase relation along the FEL pulse after the
first crystal. The artificial, linear chirp has been subtracted.
(Right plot) An enlargement of the left plot. One hundred
single-shot realizations are shown in the plots.} \label{phasetrue1}
\end{figure}
It is interesting to study the phase of the field after seeding, and
to discuss possible phase effects due to the Diamond crystal. Once
can see from Fig. \ref{TTT}, right plot, that the phase introduced
by the crystal is almost flat around the maxima of a given
oscillation, and it varies only of a fraction of a radian around the
seeding area in the low charge mode of operation\footnote{In the
long bunch mode of operation, variation in phase is much more
important, in the order of $10$ rad, but still around the value
$2\pi$, which allows for nearly Fourier limited pulses.}. This
allows for nearly Fourier limited pulses, which are characterized by
the smallest possible time-bandwidth product. We checked that this
property is inherited by the electric field after the crystal. Fig.
\ref{phaseT} shows the phase after the crystal as a function of the
position inside the FEL pulse. The large linear chirp in the seeding
part is due to the definition of the reference carrier frequency by
Genesis, which coincides with the resonant frequency of the
undulator. Since the bandstop is shifted with respect to this
reference carrier, a large, artificial linear chirp is included by
Genesis, which can be removed with appropriate subtraction. Fig.
\ref{phasetrue1} shows the phase relation along the FEL pulse after
subtraction of the linear chirp. The phase of the field varies well
below a radian, and is consistent with the analysis of Fig.
\ref{TTT}.

While the radiation is sent through the crystals, the electron beam
passes through a magnetic chicane, which accomplishes three tasks:
it creates an offset for the crystals installation, it removes the
electron microbunching produced in the first undulator, and it acts
as a delay line for the implementation of a temporal windowing
process. In this process, the magnetic chicane shifts the electron
bunch on top of the monochromatic tail created by the bandstop
filter thus temporally selecting a part of it (see the highlighted
area in Fig. \ref{Seed1}, right plot). By this, the electron bunch
is seeded with a radiation pulse characterized by a bandwidth much
narrower than the natural FEL bandwidth. For the hard X-ray
wavelength range, a small dispersive strength $R_{56}$ in the order
of a few microns is sufficient to remove the microbunching generated
in the first undulator part.  As a result, the choice of the
strength of the magnetic chicane only depends on the delay that one
wants to introduce between electron bunch  and radiation. The
optimal value amounts to $R_{56} \simeq 12 ~\mu$m for the low-charge
mode of operation. Such dispersion strength is small enough to be
generated by a short ($5$ m-long) magnetic chicane to be installed
in place of a single undulator module. Such chicane is, at the same
time, strong enough to create a sufficiently large transverse offset
for installing the crystal.

\subsection{Second radiator and second crystal}

\begin{figure}[tb]
\includegraphics[width=0.5\textwidth]{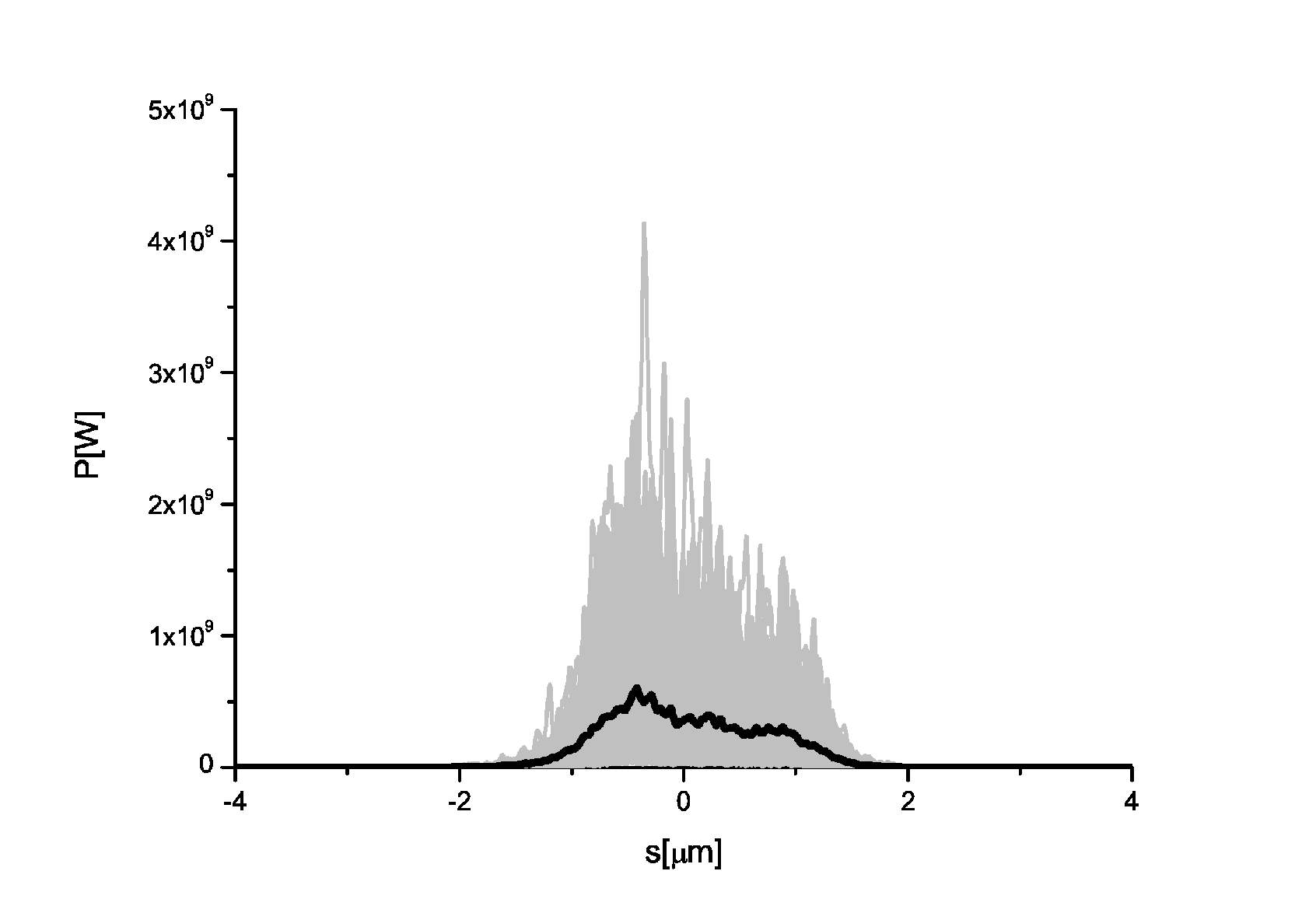}
\includegraphics[width=0.5\textwidth]{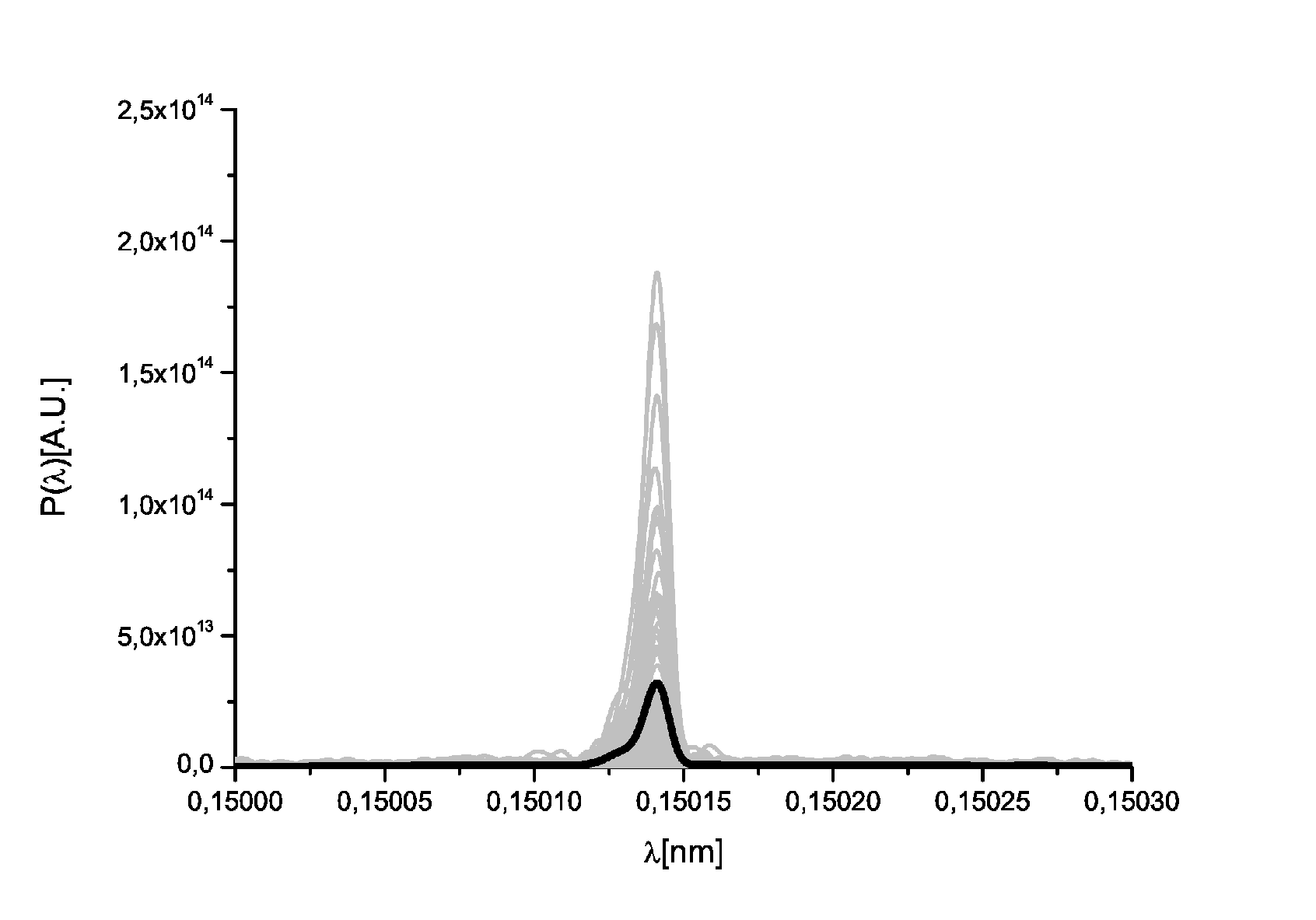}
\caption{(Left plot) Input power before the second crystal. (Right
plot) Input spectrum before the second crystal. Grey lines refer to
single shot realizations, the black line refers to the average over
a hundred realizations.} \label{seedin2}
\end{figure}

The seed signal is amplified in the second undulator, Fig.
\ref{SASE2_2}, and filtered again. The input pulse impinging on the
second crystal is shown in Fig. \ref{seedin2} in terms of power
(left plot) and spectrum (right plot).

As discussed above, due to the shortness of the first undulator, the
signal-to-noise ratio at the entrance of the second undulator cannot
be much larger than unity. Nevertheless, as shown in \cite{OURY2}
the monochromatic field amplitude at the entrance of the third
undulator will be much larger than that at the entrance of the
second.

\begin{figure}[tb]
\includegraphics[width=0.5\textwidth]{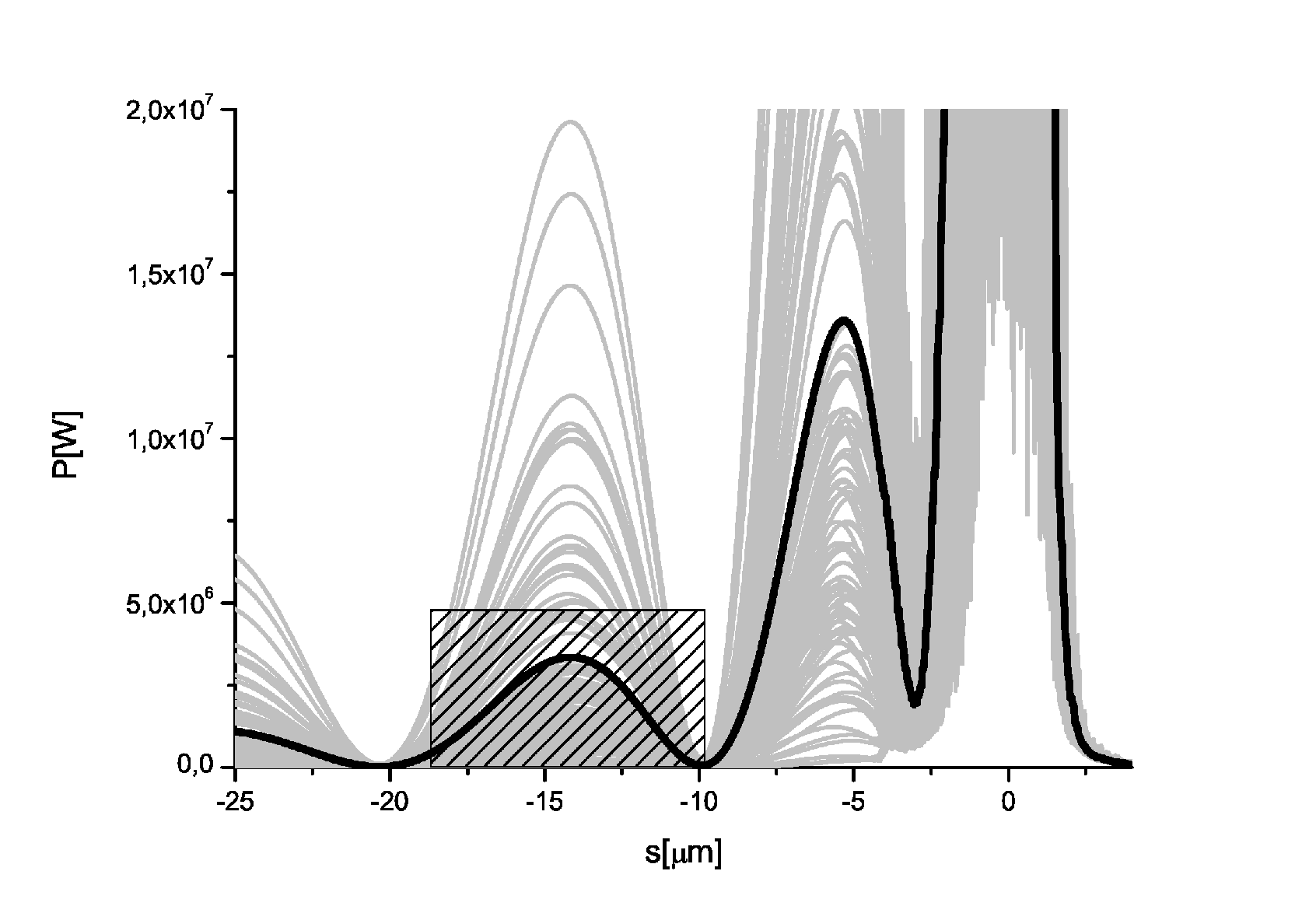}
\includegraphics[width=0.5\textwidth]{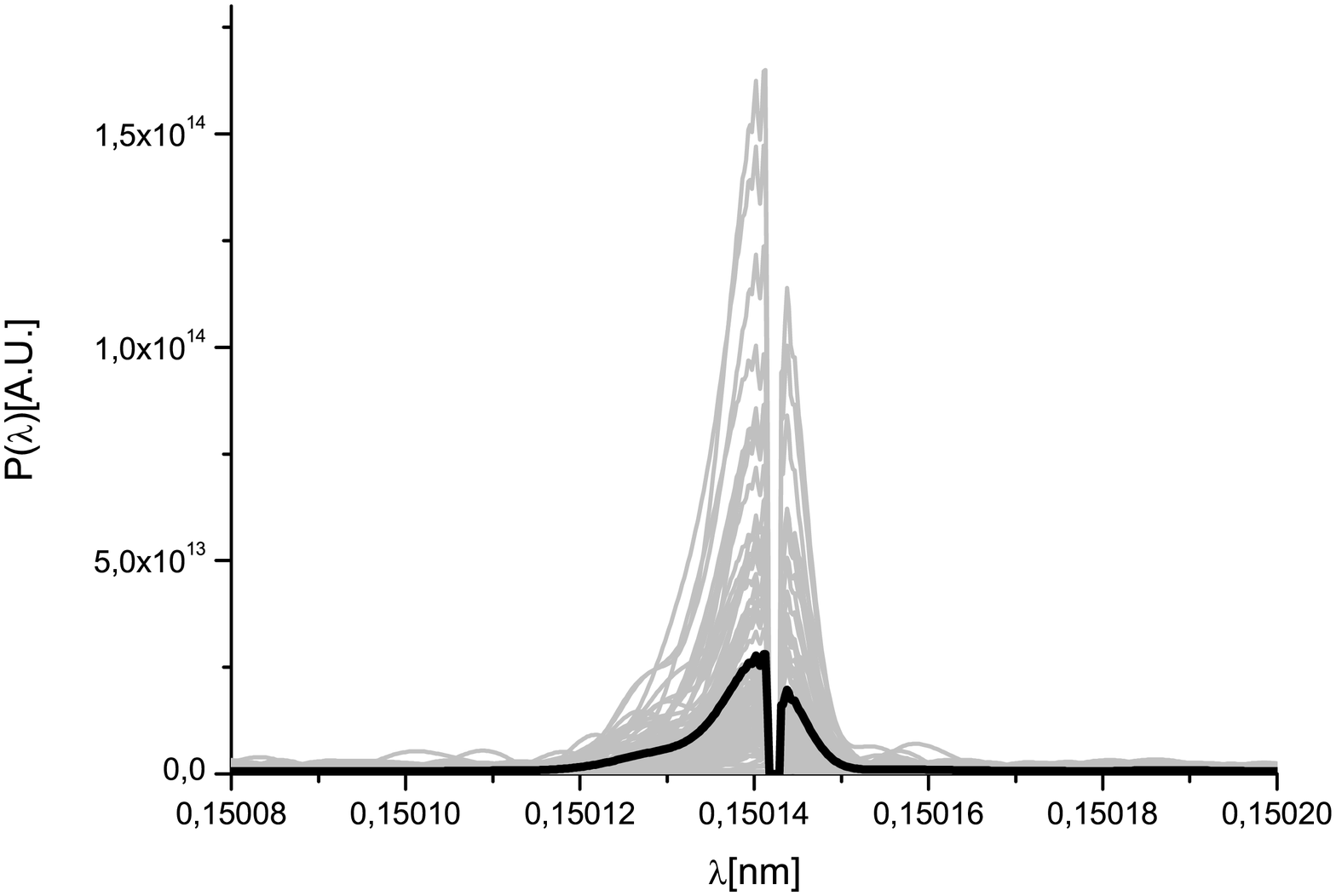}
\caption{(Left plot) Seed power after the second crystal. (Right
plot) The effect of the second crystal on the input spectrum. Grey
lines refer to single shot realizations, the black line refers to
the average over a hundred realizations.} \label{Seed2}
\end{figure}

In fact, the seed pulse after the second crystal is characterized in
power and spectrum in Fig. \ref{Seed2}. By inspection, the reader
can conclude that the average seed power at the entrance of the
output undulator is about $2$ MW, while the effective shot noise
power is about $3$ kW. A comparison with the first cascade shows
that the effective shot noise power is the same, but the
monochromatic seed power is significantly smaller i.e signal to
noise ratio at the entrance of output undulator is much larger in
the case of two cascades. compare with the case of one
amplification-monochromatization cascade. This fact can be
qualitatively explained as follows.

After amplification in the second undulator (Fig. \ref{seedin2}),
the bandwidth of the FEL pulse related with the monochromatic signal
that impinges on the second crystal is near to the transform-limited
bandwidth of the electron bunch i.e. $c/\sigma_e$. As a result,
assuming the same amplification in the two cascades, the signal to
noise ratio is enhanced by a factor $\sigma_e \Delta
\omega_{SASE}/c$, where $\Delta \omega_{SASE}$ is the SASE
bandwidth. A rough estimate for the signal to noise ratio at the
entrance of the third (output) undulator is therefore\footnote{From
our reasoning we conclude that the enhance of the signal-to-noise
ratio is about a factor $100$. This is the result of two effect. the
first is the before-mentioned amplification of the monochromatic
seed in the first cascade i.e. $\sigma_e \Delta \omega_{SASE} \sim
10$. The second reason is the presence of an extra cell in the
second cascade, which additionally enhance the signal-to-noise ratio
of a factor $\sim 10$. Of course, a configuration with $6$ cells
followed by $5$ cells will lead to the same monochromatic seed at
the entrance of the output undulator as the configuration with $5$
cells followed by $6$ cells.} $P_\mathrm{seed2}/P_\mathrm{n} \sim
(P_\mathrm{seed1}/P_\mathrm{n}) \sigma_e \Delta \omega_{SASE}/c \gg
1$. Since this value is much larger than unity, we conclude that a
double cascade self-seeding scheme using a single-crystal
monochromator is insensitive to noise and non-ideal effects.


\subsection{Output characteristics}

\begin{figure}[tb]
\includegraphics[width=0.5\textwidth]{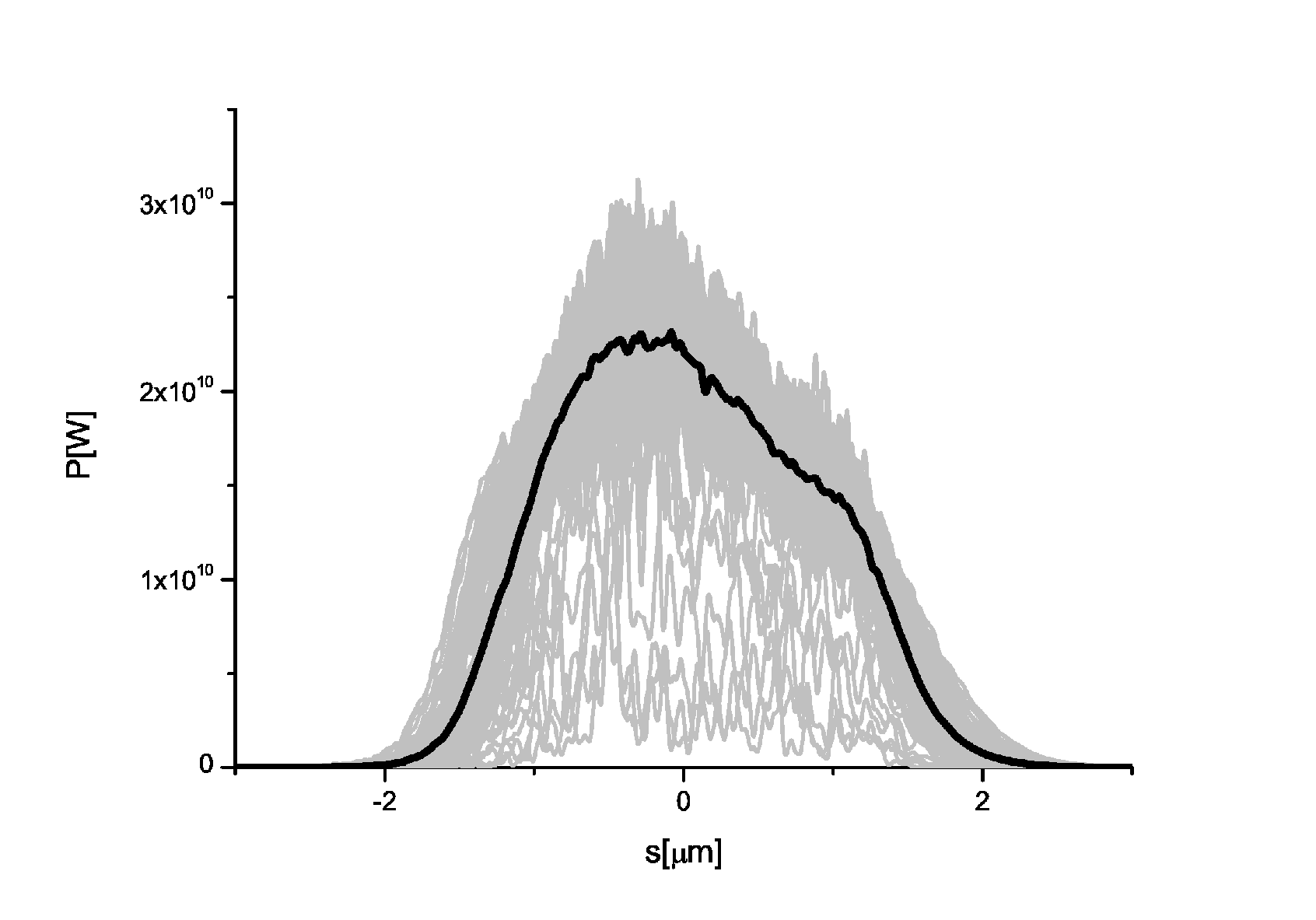}
\includegraphics[width=0.5\textwidth]{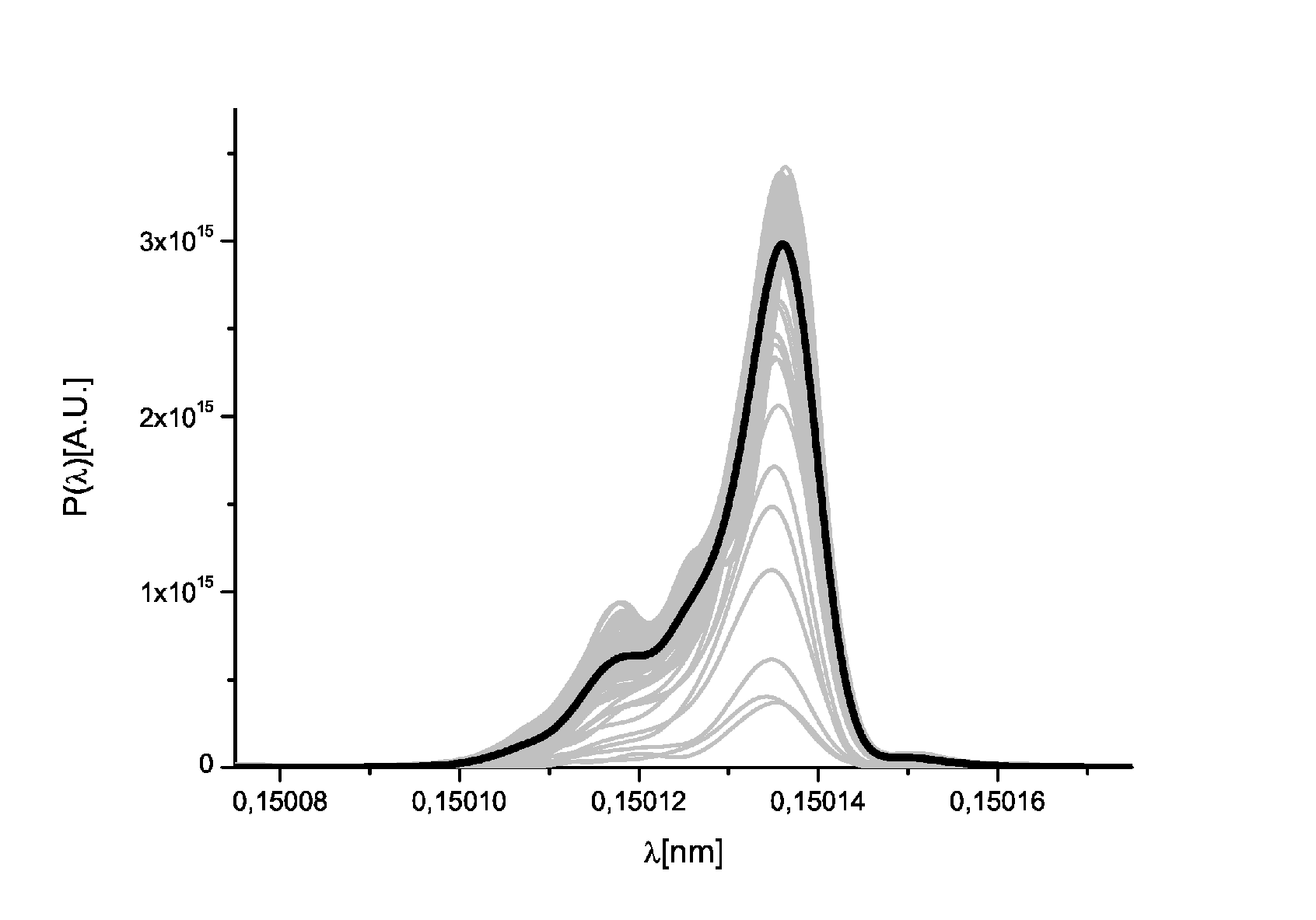}
\caption{Power distribution  and spectrum of the X-ray radiation
pulse at saturation without tapering. Here the output undulator is 7
cells long. Grey lines refer to single shot realizations, the black
line refers to the average over a hundred realizations.}
\label{Outnotap}
\end{figure}
\begin{figure}[tb]
\includegraphics[width=0.5\textwidth]{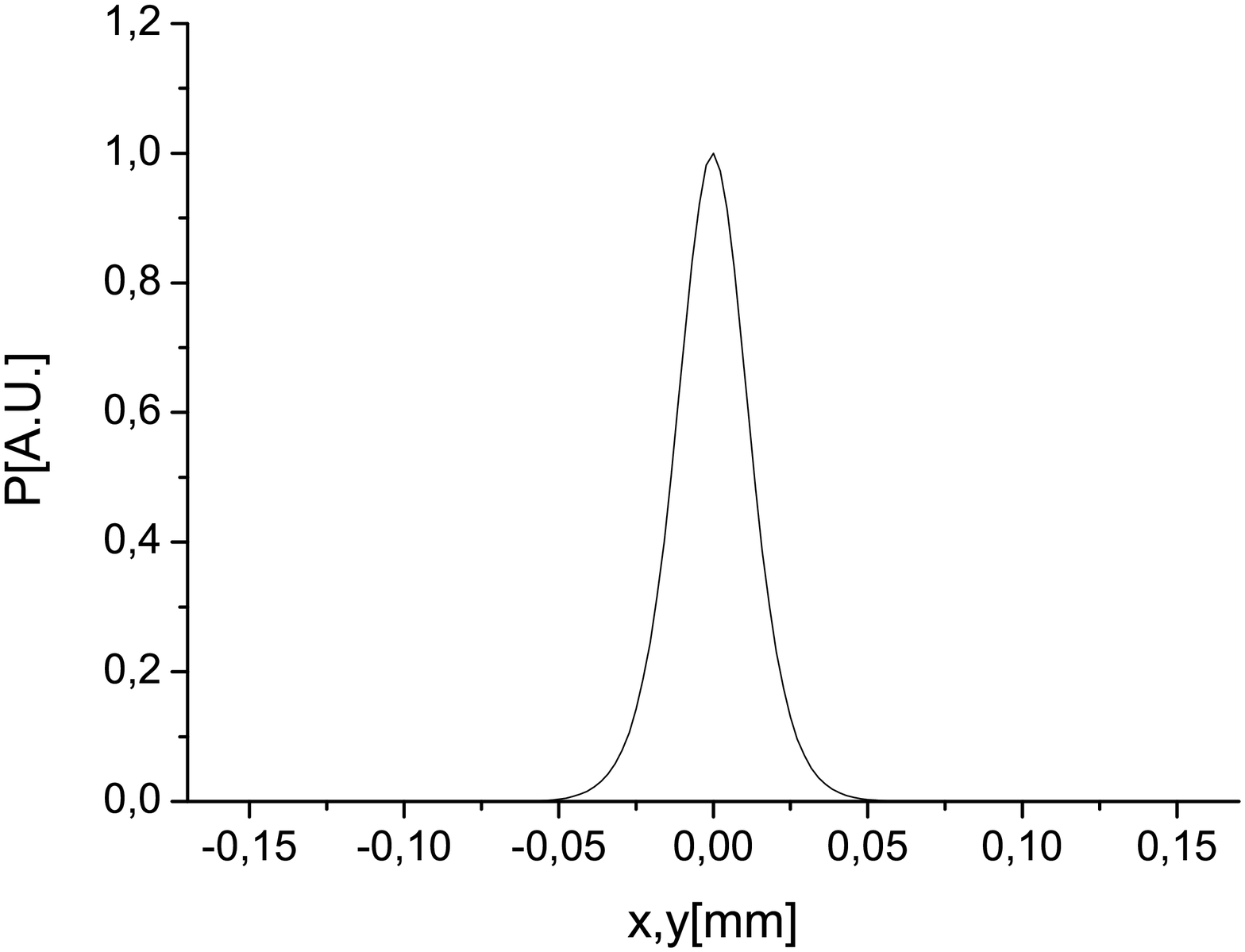}
\includegraphics[width=0.5\textwidth]{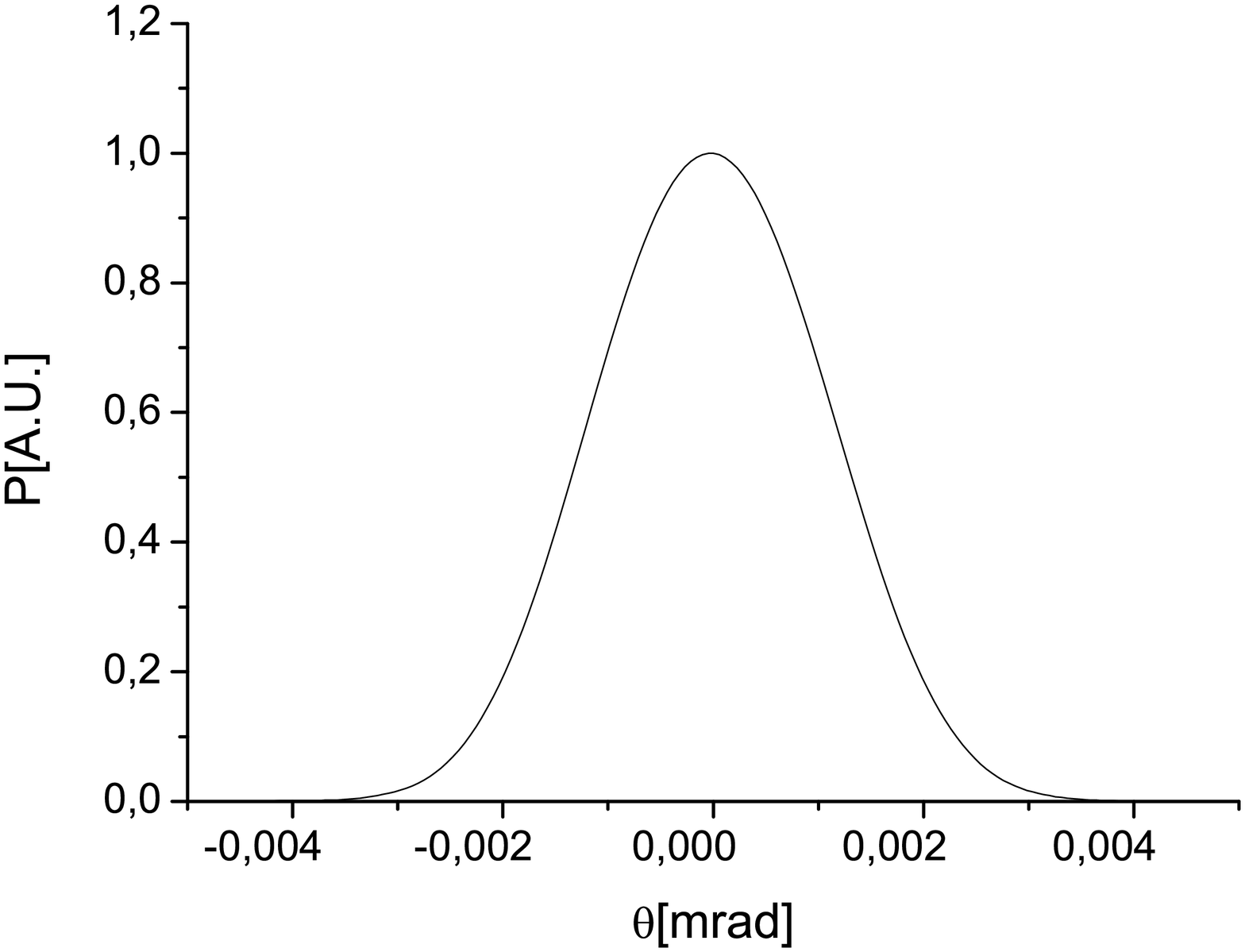}
\caption{(Left plot) Transverse radiation distribution at saturation
without tapering at the exit of the output undulator. (Right plot)
Directivity diagram of the radiation distribution at saturation
without tapering at the exit of the output undulator.}
\label{outsatspot}
\end{figure}
%

After the seed leaves the second crystal, it is superimposed to the
electrons at the entrance of the last undulator, where it is
amplified up to saturation,  Fig. \ref{Outnotap}. One finds an
almost bandwidth-limited output pulse, i.e. a pulse with no
frequency modulation. The output power is in excess of $20$ GW, and
the nearly Fourier-limited bandwidth corresponds to a time-bandwidth
product $\Delta t \cdot \Delta \omega \simeq 6.6$, to be compared
with the minimal time-bandwidth product for a Gaussian\footnote{The
time-bandwidth product constitutes a figure of merit for qualitative
analysis only. In fact, in our case the time-domain pulse is
non-Gaussian and non-symmetric. The time-bandwidth limit for a
stepped-profile pulse is, for example, about $1.35$ times larger
than that for a Gaussian pulse. The comparison with the Gaussian
case qualitatively shows that we are not far from the ultimate
time-bandwidth product, and the difference can be explained, at
least partially, by the asymmetry of the pulse profile in the time
domain. } given by $2.8$. The transverse radiation distribution and
divergence at the exit of the output undulator are shown in Fig.
\ref{outsatspot}.

\begin{figure}[tb]
\includegraphics[width=1.0\textwidth]{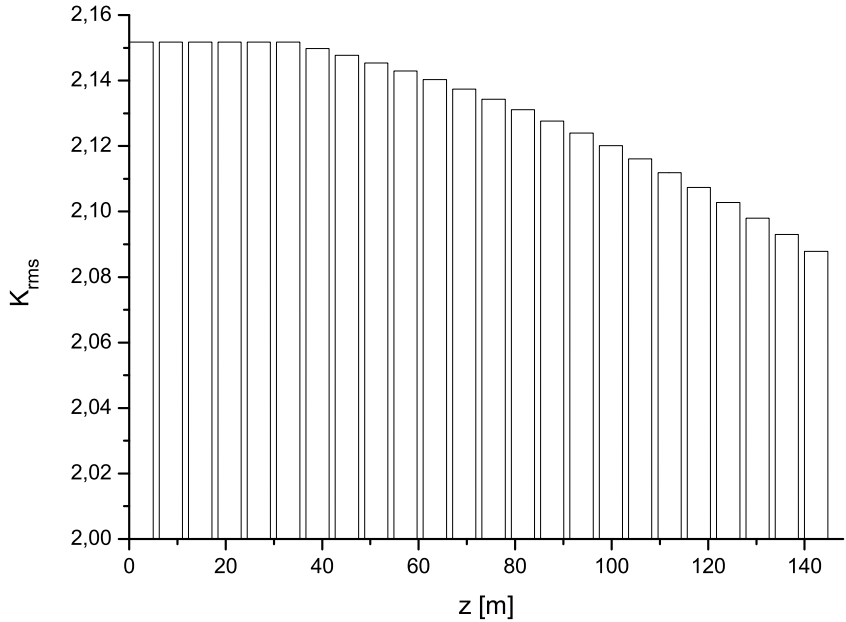}
\caption{Taper configuration for high-power mode of operation at
$0.15$ nm.} \label{Krms}
\end{figure}
\begin{figure}[tb]
\includegraphics[width=0.5\textwidth]{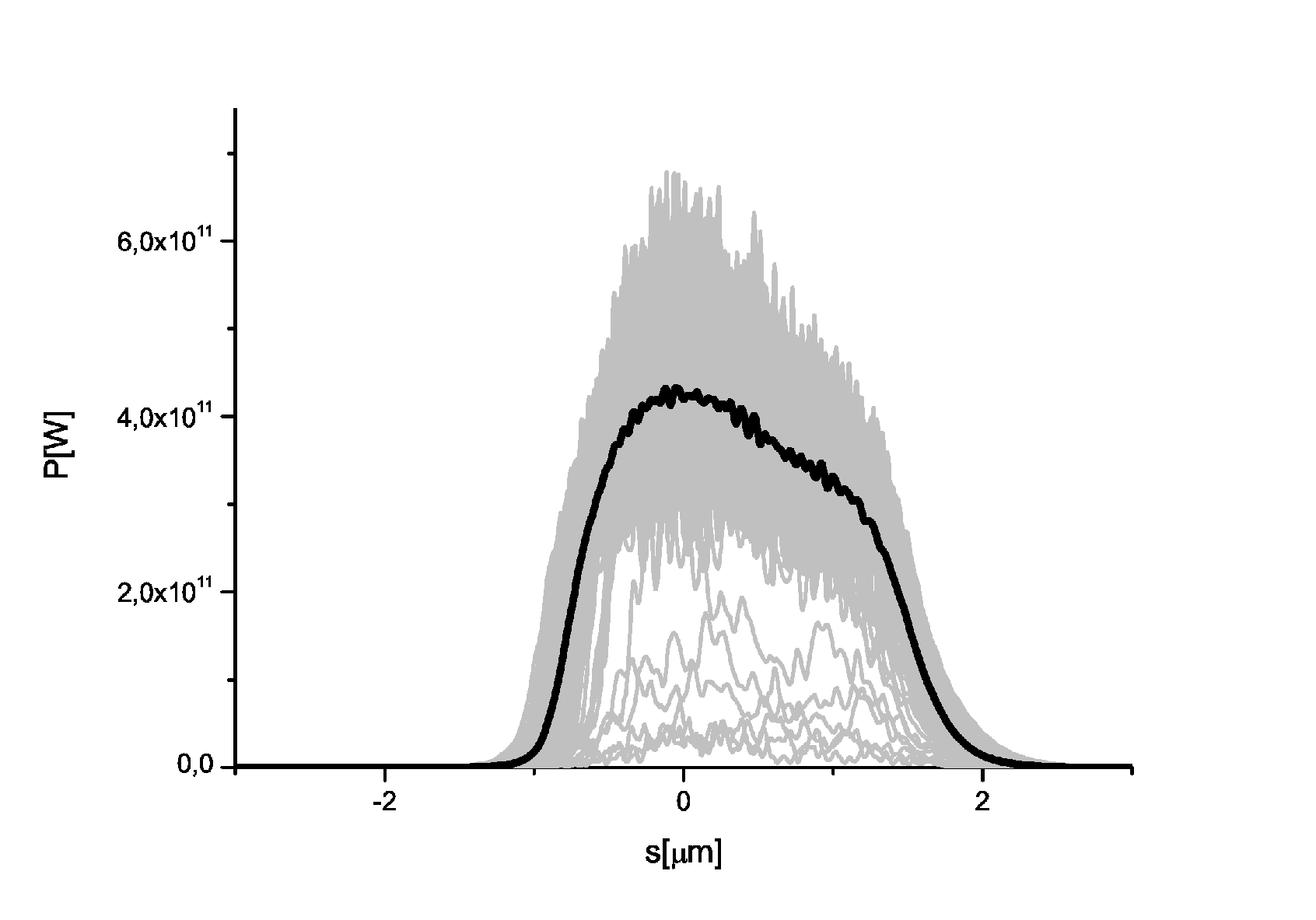}
\includegraphics[width=0.5\textwidth]{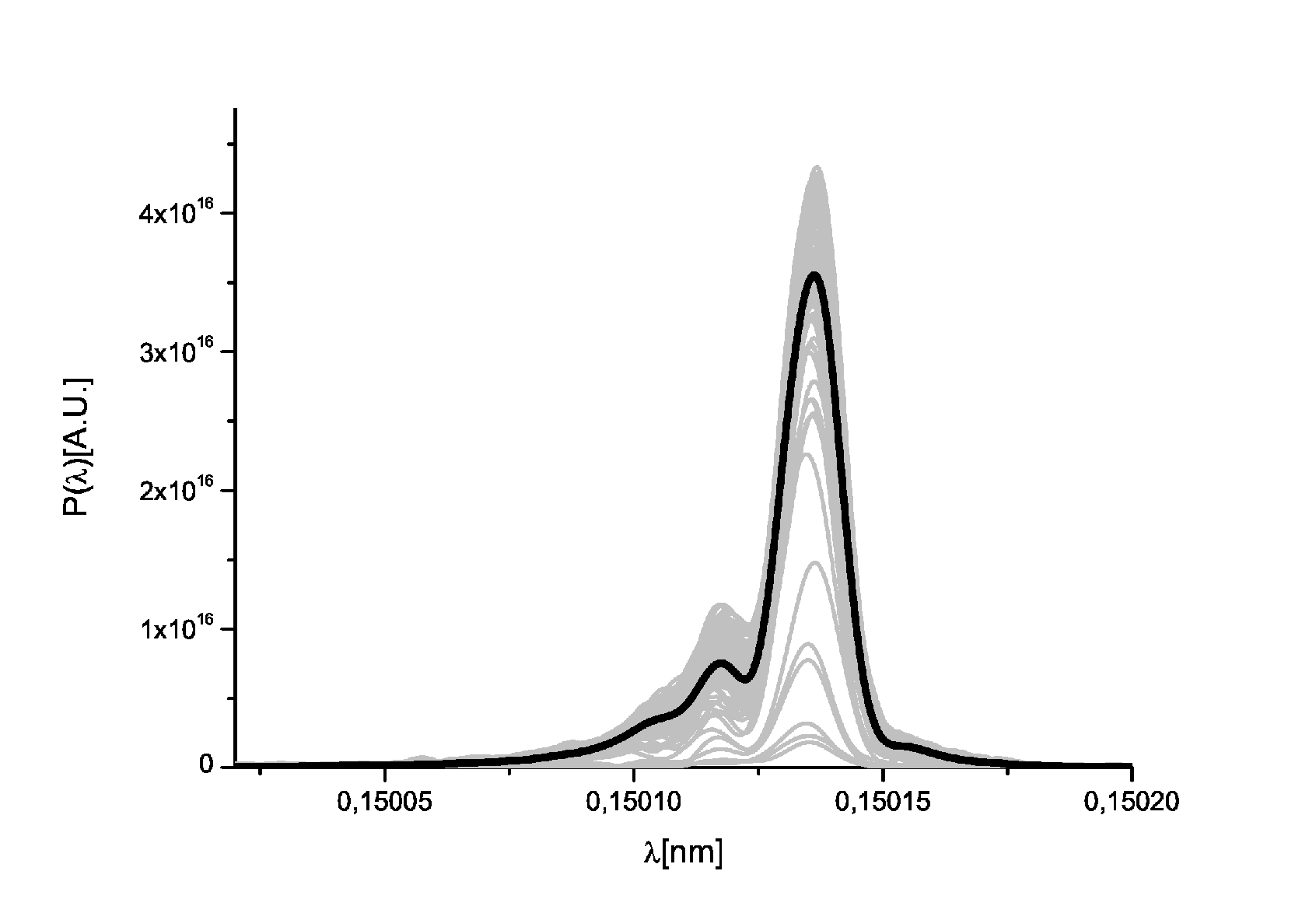}
\caption{(Left plot) Output power at saturation in the case of
tapering. (Right plot) Output spectrum at saturation in the case of
tapering. Grey lines refer to single shot realizations, the black
line refers to the average over a hundred realizations.}
\label{Outtap}
\end{figure}
\begin{figure}[tb]
\includegraphics[width=0.5\textwidth]{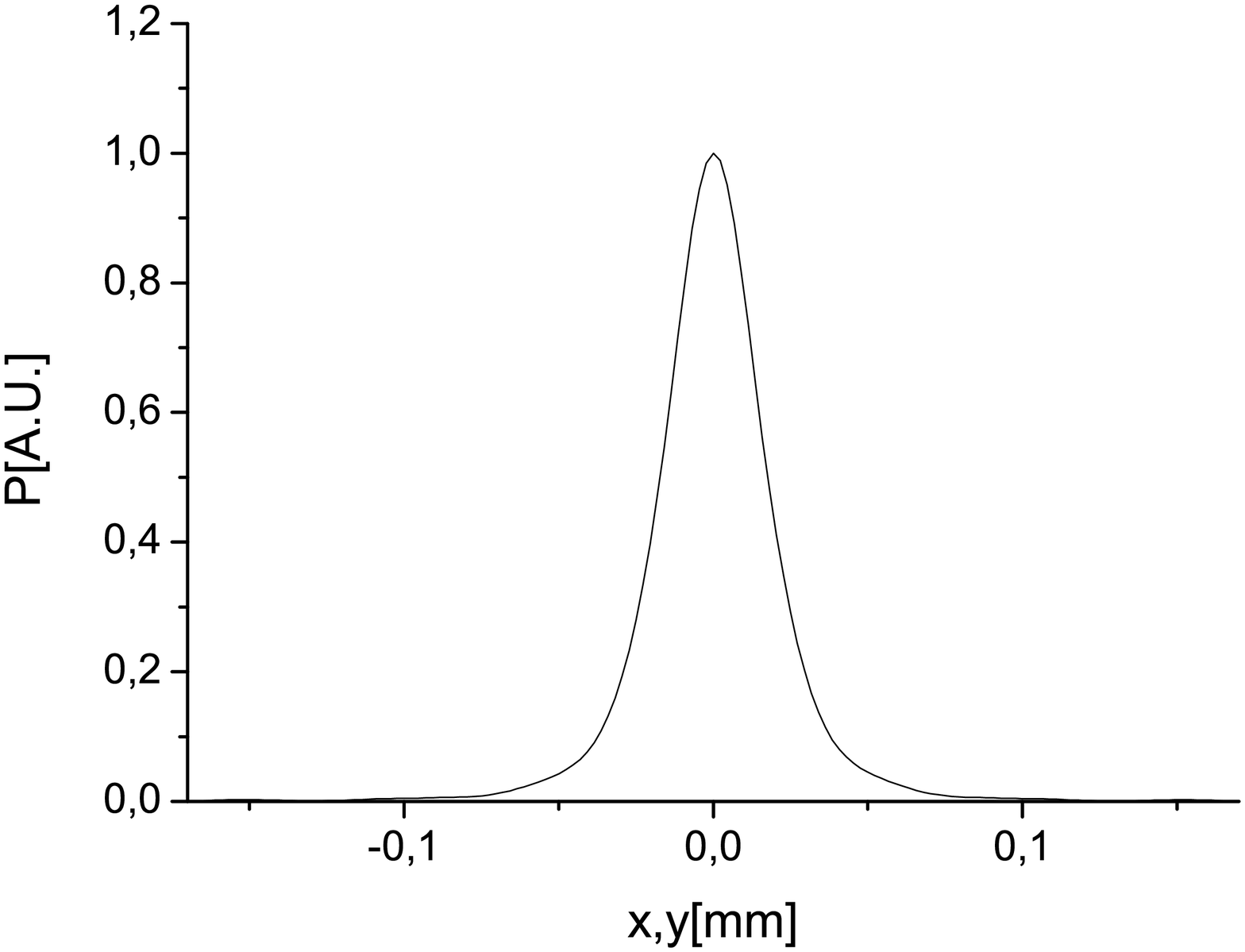}
\includegraphics[width=0.5\textwidth]{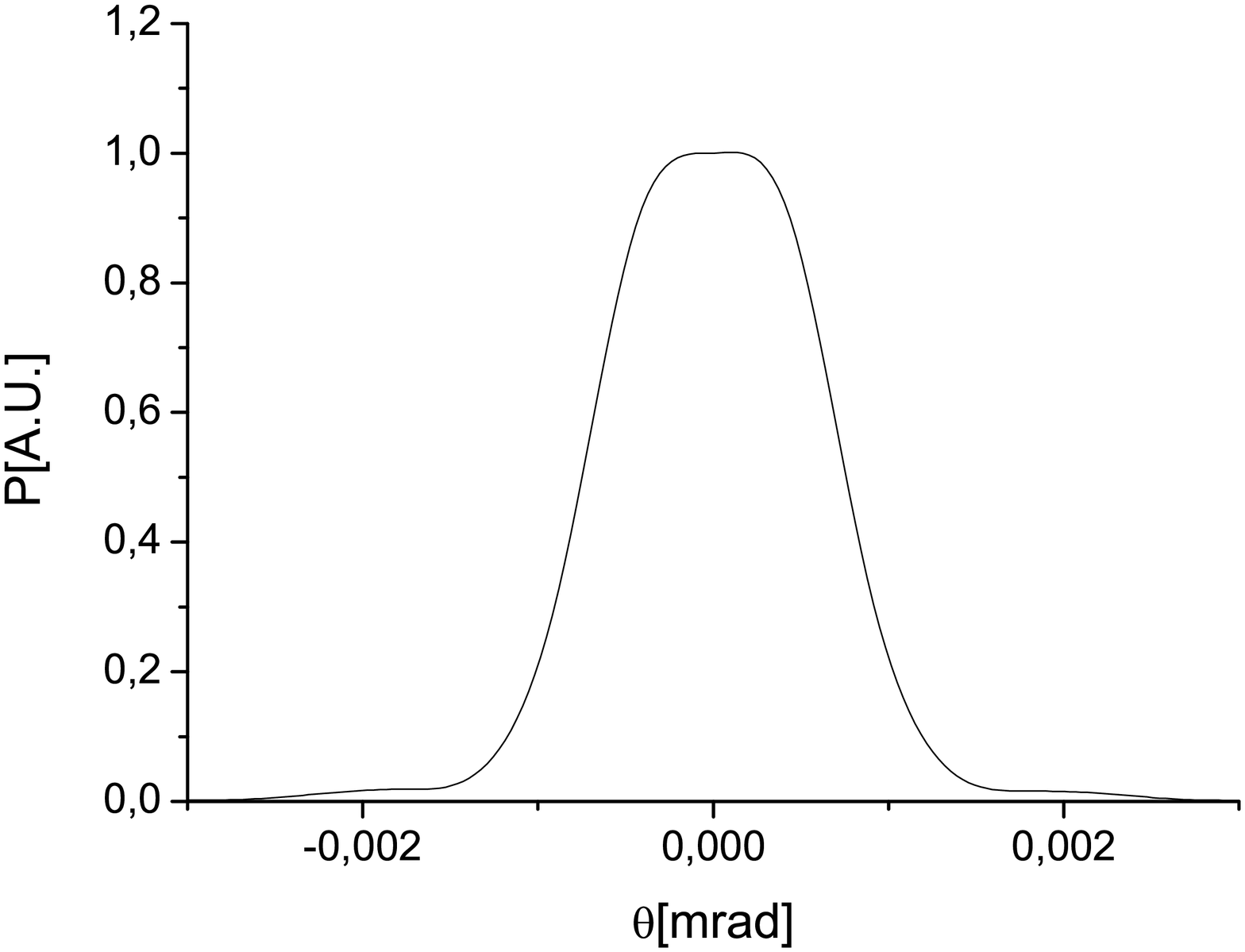}
\caption{(Left plot) Transverse radiation distribution in the case
of tapering at the exit of the output undulator. (Right plot)
Directivity diagram of the radiation distribution in the case of
tapering at the exit of the output undulator.} \label{outtapspot}
\end{figure}

\begin{figure}[tb]
\includegraphics[width=1.0\textwidth]{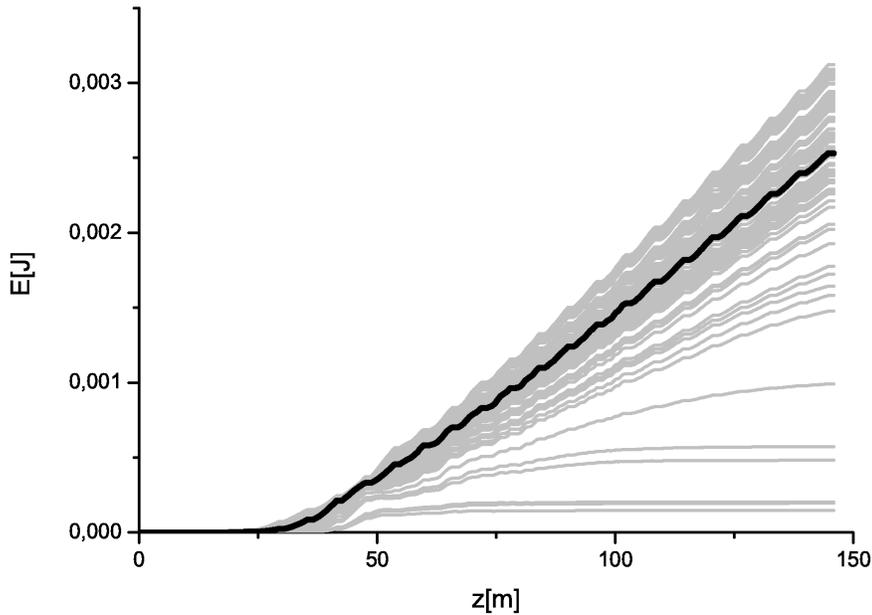}
\caption{Energy in the X-ray radiation pulse versus the length of
the output undulator. Tapering is considered according to the law in
Fig. \ref{Krms}. Grey lines refer to single shot realizations, the
black line refers to the average over a hundred realizations.}
\label{Eout}
\end{figure}
One can prolong the exchange of energy to the photon beam by
tapering the last part of the radiator on a segment-to-segment
basis. The optimal tapering law is found empirically and is
shown\footnote{Note that additional tapering should be considered to
keep the undulator tuned in the presence of energy loss from
spontaneous radiation. In Fig. \ref{Krms} we present a tapering
configuration which is to be considered as an addition to this
energy-loss compensation tapering. } in Fig. \ref{Krms}. Taper
begins from the first six cells. The output power and spectra at the
exit of the setup are shown in Fig. \ref{Outtap}. A final output
power in the TW level ($400$ GW) can be reached with the help of the
baseline undulator, while the radiation pulse remains nearly Fourier
limited, with a time-bandwidth product $\Delta t \cdot \Delta \omega
\simeq 8.0$. The transverse radiation distribution and divergence at
the exit of the output undulator are shown in Fig. \ref{outtapspot}.
The evolution of the energy in the radiation pulse as a function of
the output undulator length is shown in Fig. \ref{Eout}.

To conclude, it is interesting to compare our previous calculations
for the European XFEL \cite{OURY3} with the ones discussed in the
present article. In fact, in this work we make use of the phase
space distribution from start-to-end simulations, we include wakes
and, additionally, we reduced the energy of electron beam (of $3.5$
GeV) and shortened the undulator length (of $7$ cells). The
undulator period is also slightly different ($40$ mm, compared with
$48$ mm in the previous study). As a result of these changes, at
first glance we should expect some performance degradation. The
simulation results presented in this section show that this is not
the case. The reduction of energy and undulator length can only lead
to performance degradation. Similar reasoning holds for the presence
of wakes. However, performance improvement is to be expected due to
a better slice emittance. In previous calculations we considered
$0.4~\mu$m, uniformly distributed normalized emittance based on the
experience of the LCLS. In the present article, based on start to
end simulations for the European XFEL we consider a slice emittance
approximately two times smaller ( $< 0.25 ~\mu$m ) within all bunch
length for both directions (see Fig. \ref{s2E}). As our simulations
have shown, this leads to performance increase and compensates for
detrimental factors.

\section{Conclusions}

In the baseline SASE mode, the European XFEL will provide
transversely coherent beams but only limited longitudinal coherence.
However, an important goal for advanced XFEL sources is the creation
of transform-limited pulses of radiation, which are important for
several reasons. First, they naturally provide the maximum intensity
within the minimum photon energy window for given pulse length, so
that no monochromator is needed in the experimental hall or at the
photon beam lines. Second, when the radiation beam is
monochromatized down to the Fourier transform limit, a variety of
very different techniques become feasible, leading to further
improvement of the XFEL performance. For example, as we demonstrated
in \cite{OURY3} for an ideal electron beam,  monochromatization
effectively allows one to use an undulator tapering technique
enabling a high power mode of operation. Such mode of operation is
highly desirable. In fact, in spite of the unprecedented increase in
peak power of X-ray pulses from SASE XFELs, compared with third
generation facilities, some applications require still higher photon
flux. In the present paper we extend our consideration to a
realistic electron beam distribution at the undulator entrance. In
particular we propose a study of the performance of a cascade
self-seeding scheme with single crystal monochromators for European
XFEL, based on start-to-end simulations and accounting for undulator
wakefields \cite{S2ER}. By combining the two techniques of cascade
self-seeding and undulator tapering, we find that TW-level X-ray
pulses can be generated with minimal modifications to the baseline
mode of operation. The results of our analysis indicate that our
self-seeding scheme is not significantly affected by non-ideal
electron phase space distribution and has about the same performance
in the start to end case as for the electron beam with ideal
properties.

\section{Acknowledgements}

We are grateful to Massimo Altarelli, Reinhard Brinkmann,
Serguei Molodtsov and Edgar Weckert for their support and their interest during the compilation of this work.

\end{document}